\documentclass[preprint,12pt]{elsarticle}

\usepackage{amssymb}
\usepackage{url}
\usepackage{graphicx}
\usepackage{subfig}
\usepackage{comment}
\usepackage{array}
\usepackage{booktabs}
\usepackage{multirow}
\usepackage{adjustbox}
\usepackage{color}
\usepackage{xcolor}
\usepackage{hyperref}
\usepackage{makecell}
\usepackage{longtable}

\begin{document}

\begin{frontmatter}


 \author[a]{Mahsa Tavakoli\corref{cor1}}
 \ead{mtavako5@uwo.ca}
 \cortext[cor1]{Corresponding Author}
 

\title{Multi-Modal Deep Learning for Credit Rating Prediction Using Text and Numerical Data Streams
}
\author[Rohit]{Rohitash~Chandra}
\ead{rohitash.chandra@unsw.edu.au}
\author[a]{Fengrui~Tian}
\ead{ftian9@uwo.ca}
\author[a]{Cristi\'{a}n~Bravo}
\ead{cbravoro@uwo.ca}

\affiliation[a]{organization={Department of Statistics and Actuarial Science, University of Western Ontario},
            addressline={}, 
            city={London},
            postcode={N6A 5B7}, 
            state={Ontario},
            country={Canada}}

\affiliation[Rohit]{organization={
Transitional Artificial Intelligence Research Group, School of Mathematics and Statistics, UNSW Sydney},
            state={Sydney 2052},
            country={Australia}}

\begin{abstract}
Knowing which factors are significant in credit rating assessments leads to better decision-making. However, the focus of the literature thus far has been mostly on structured data, and fewer studies have addressed unstructured or multimodal datasets. In this paper, we present an analysis of the most effective architectures for the fusion of deep learning models to predict company credit rating classes, using structured and unstructured datasets of different types. In these models, we tested various combinations of fusion strategies with selected deep-learning models, including convolutional neural networks (CNNs) and variants of recurrent neural networks (RNNs), and  pre-trained language models (BERT). We study data fusion strategies in terms of level (including early and intermediate fusion) and techniques (including concatenation and cross-attention). Our results show that a CNN-based multi-modal model with a hybrid fusion strategy outperformed other multimodal techniques. In addition, by comparing simple architectures with more complex ones, we found that more sophisticated deep learning models do not necessarily produce the highest performance. 
Furthermore, we found that the text channel plays a more significant role than numeric data, with the contribution of text achieving an AUC of 0.91, while the maximum AUC of numeric channels was 0.808.
Finally,   rating agencies on short, medium, and long-term performance show that Moody's credit ratings outperform those of other agencies like Standard \& Poor's and Fitch Ratings.
\end{abstract}




\end{frontmatter}


\section{Introduction}


The issue of credit rating prediction has received interest from many researchers due to its profit potential in the financial market \cite{loffler2020systemic}, including portfolio risk management \cite{yijun2009artificial}. This interest has been driven by the importance of credit scoring for accurate assessment of credit risk, which is crucial for banks and other financial institutions to make informed decisions regarding lending money to corporations \cite{he2010domain}. In addition, credit scoring is important for predicting the probability of default of a corporate borrower, which is critical to properly assess risk for what are usually large loans \cite{liu2022rmt}. Related studies have focused on two major approaches, namely linear and non-linear models. Most credit scoring models in practice are deployed using logistic regression \cite{thomas2017credit}. In addition to this traditional statistical model, machine learning models such as k-nearest neighbour, decision trees, random forests, support vector machines, and multilayer perception (MLPs, or simple neural networks) have been employed for credit rating\cite{dastile2020statistical,CreditScoringBenchmark2024}, as well as in other domains in financial applications \cite{zhu2021research, nazareth2023financial, sarker2021machine}.

Studies have also explored the potential of integrating different models to form ensemble models, which are heavily applied in financial scenarios such as credit rating \cite{jain2007hybrid}. Deep learning models \cite{lecun2015deep} are prominent machine learning models that have the capacity to learn from complex data that revolutionized  domain applications \cite{deng2014tutorial} such as image processing, speech recognition \cite{{mehrish2023review}}  and natural language processing (NLP) \cite{otter2020survey}, and medicine \cite{cai2020review}. Deep learning models are also  prominent in finance and economics \cite{ozbayoglu2020deep}. Among them, Convolutional Neural Networks (CNNs) \cite{o2015introduction,feng2020every} are deep learning models designed for image and video processing tasks \cite{khodaee2022forecasting}, but are also used for time series prediction \cite{chandra2021evaluation}. 
Recurrent Neural Networks (RNNs) \cite{medsker2001recurrent} are deep learning models suited for modelling temporal sequences \cite{zhang2018spatial} which made them popular for NLP tasks \cite{chowdhary2020natural}. Some famous RNN architectures include the Long Short-Term Memory (LSTM) \cite{hochreiter1997long} networks, Gated Recurrent Units (GRU) \cite{cho2014properties}, and Transformer models \cite{vaswani2017attention}, which developed further the LSTM architecture by equipping with an attention mechanism based on cognition, among other improvements. LSTM networks have the property of remembering information over long periods of time, overcoming the problems of vanishing gradient of canonical RNNs and GRUs. Shen et al. \cite{shen2018deep} applied GRU  to predict trading signals of stocks part of the Hang Seng and the S\&P 500 indices, and obtained better performance when compared to support vector machines.

NLP tasks such as machine translation \cite{nguyen2020neural}, text classification \cite{cambria2014jumping} and sentiment analysis \cite{chandra2021covid}, involve processing raw text that features unstructured information using pre-trained deep learning models \cite{{min2023survey}}. The  Bidirectional Encoder Representations from Transformers (BERT) \cite{devlin2018bert} is a pre-trained deep learning model that overcame the limitation of unidirectional language models, and outperformed the conventional language models in key NLP tasks including machine translation, text classification, and speech processing \cite{karita2019comparative}.

Deep learning models have been extensively used in financial applications \cite{ozbayoglu2020deep} such as financial time series forecasting \cite{sezer2020financial}.   \cite{min2023recent} found that the majority of financial applications have utilized structured datasets, including time series and tabular data with numerical features \cite{wang2015survey}. However, limited studies where unstructured data such as text has been applied for financial applications have been explored \cite{li2020multimodal}. In addition to employing unstructured data, some researchers have also presented models for applying both structured and unstructured data, leading to the presentation of multimodal techniques and application \cite{stevenson2021value,sait2021deep,hu2023collaborative,hou2022multi,park2022multimodal,korangi2023transformer, zandi2024attentiongraph}. In models for unstructured data, mixing information from various channels is known as a \textit{multimodal fusion strategy}, referring to the process of combining information by integrating different channels \cite{d2009exploring}. A review by Soujanya et al. \cite{poria2017review} indicated that multimodal models are significantly more accurate than unimodal models. 

Apart from the type of fusion, the fusion level (fusion order) is another topic in multimodal research. There are three approaches in terms of fusion level: early (or signal) fusion, in which raw channels or modalities are combined at the earliest stage of the pipeline; the intermediate fusion approach, in which the modalities are integrated after the feature extraction process but before classification; and the late or decision-level fusion, in which the outputs are combined for the classification decision \cite{zhu2015fusing}. Boulahia et al. \cite{boulahia2021early} studied the impact of fusion level in the multimodal recognition process and found that the intermediate level fusion had the strongest impact compared to early and late fusion. Lee et al. \cite{lee2020multimodal} applied multimodal deep learning for stock market prediction and reported  that intermediate fusion led to better performance compared to late fusion.
Luo et al. \cite{luo1988multisensor} demonstrated that multimodal models can mix different types of modalities, often using a concatenation technique as a simple fusion strategy.

To the best of our knowledge, no research has been developed regarding the prediction of credit rating classes using various levels and types of fusion for combining different modalities. An example of a more modern type of fusion technique involves cross-attention as the core fusion strategy \cite{huang2019ccnet,hou2019cross,lee2018stacked}. 

Further extending multimodal fusion strategies, cross-attention has been recently used to replace concatenation and improve fusion performance over multi-modal data. Cross-attention is an attention-based mechanism that helps to identify the correlation and interaction of the different modalities of data when combining them \cite{meng2020survey}. Few studies have investigated the effect of unstructured or a combination of unstructured and structured data on credit rating prediction. Moreover, due to the complexity of the analytical process for unstructured data in comparison to structured data, only recent research has investigated its impact on many applications including financial prediction \cite{pejic2019text}.

In this paper, we  apply multimodal deep learning models for  company credit rating prediction using structured and unstructured datasets of different types. The structured datasets include market, bond, financial ratios, and related information (agency type and past rating), while the unstructured dataset contains text documents of earnings call transcripts. The earnings call is a periodic conference led by the executives of companies to compile company performance and present it to the market \cite{frankel2017using}. We investigate the effect of different multimodal data fusion strategies, including concatenation and cross-attention. Besides the fusion type, we also investigate the impact of the order (level) of fusion in a model on performance. Different combinations of fusion type (concatenation and cross-attention) and order (early or intermediate), could lead to four possible modes for the model’s structure. We develop each possible mode (channel) in a multimodal data fusion strategy, based on four commonly used deep learning models (CNN, LSTM, GRU and BERT). Therefore, we provide a comprehensive evaluation of  the combination of 16 models into four major categories in terms of fusion type and fusion level. Moreover, by activating and inactivating each channel in the best model, the value of each channel of structured data and text modality, further helping to evaluate the usefulness of unstructured data. We also provide a comparison of the impact of the different rating agencies on the short, medium, and long-term performance of the model.


To finalize our analysis, and in light of the drastic effect  the COVID-19 pandemic had on the economy and financial markets, plus its disruption of international trade \cite{brodeur2021literature}, we have included its impact on our models. 
Zeren et al. \cite{zeren2020impact} indicated that investing in the stock market may not have been the correct option for investors during the peak of the pandemic due to emerging unpredictable conditions. A more appropriate option might have been bonds, as research has shown that bonds offered diversification during the COVID-19 crisis. For instance, Papadamou et al. \cite{papadamou2021flight} found that stock and bond returns moved in opposite directions during the pandemic, demonstrating a "flight-to-quality" where investors sought the relative safety of bonds, especially when stock markets were highly volatile.
Therefore, we need to investigate the performance of our multimodal deep learning models during the COVID-19 pandemic and compare with results prior to the pandemic. Since our data overlaps the COVID-19 pandemic period, we also evaluate the models by considering fluctuations in the market during the pandemic period. 

The remainder of this work is as follows: In Section 2, we review  related literature and outline the methodology in Section 3  for structuring various multimodal models. Section 4 presents the results highlighting model comparison.  Section 5 presents a discussion of the results and finally, we conclude the study in Section 6.

\section{Related Works}
\label{sec:related works}

\subsection{Deep learning for financial prediction}

Lately, deep learning models have been applied to many financial prediction problems based on structured data, including early delinquency \cite{chen2021predicting}, profit scoring \cite{fitzpatrick2021can,lappas2021machine,dastile2020statistical,oskarsdottir2019value}, credit scoring \cite{gunnarsson2021deep,kvamme2018predicting}, price movement prediction \cite{jabeur2021forecasting,zhao2023deep}, trader classification \cite{kim2020can}, fraud detection \cite{seera2021intelligent}, and many others. In a survey, Huang et al. \cite{huang2020deep} compared a wide range of models in the application of finance and banking domains including stock market prediction, banking default risk, and credit rating. They considered deep models with machine learning models and found that CNN and LSTM-based models provided the best performance in most cases. Neagop et al. \cite{neagoe2018deep} used CNN and multilayer perceptron (MLP) for credit scoring and found that CNN achieved superior performance. Golbayani et al. \cite{golbayani2020application} found that LSTM models outperformed CNNs in predicting corporate credit ratings issued by the Standard \& Poor's (S\&P) rating agency.

Moreover, Qian et al. \cite{qian2023soft} developed a CNN architecture with a soft reordering mechanism that could adaptively reorganize tabular data for better learning, that outperformed related models for credit scoring. Adisa et al. \cite{adisa2022credit} explored the effectiveness of combining multiple classifiers into an ensemble for credit scoring prediction.  In the domain of credit risk, Korangi et al. \cite{korangi2022transformer, zandi2024attentiongraph} applied a transformer-based model  for panel-data classification  for midcap companies to detect complex, non-linear relationships over a long range of time series.
Huang et al. (2020)\cite{huang2020deep} provide a thorough review of deep learning applications in finance and banking, focusing on model preprocessing, input data, and evaluation techniques. Their study highlights state-of-the-art methods and their impact on improving accuracy in classification tasks within the financial domain, offering insights into advanced model implementations for better financial decision-making.

In recent advancements, industry-specific challenges have been utilized using deep learning. Cheng et al. \cite{cheng2024multi} presented a novel approach for predicting business processes in call centres, utilizing a multimodal fusion of data based on customer interactions and demonstrated enhanced prediction accuracy and multitasking capabilities.
Tan et al. \cite{tan2024asset} introduced a deep learning framework that amalgamated structured market data with unstructured media content for predicting stock movements using a Context-Aware Hierarchical Attention Mechanism that improved the interpretability of predictions by providing visual clues. 
Che et al. \cite{che2024predicting}  developed  a deep learning approach that leveraged financial indicators, current reports, and interfirm networks to predict financial distress using modality-specific attention mechanisms that outperformed traditional models. However, the approach faced challenges due to high computational demands and dependence on the quality of multimodal data.
 Zhang et al. \cite{zhang2024smpdf} utilized a deep learning model that integrated stock price data with textual information through bilinear pooling and captured both quantitative and qualitative market indicators using Bidirectional-LSTM and self-attention mechanism. 

\subsection{NLP for finance problems}
The datasets commonly used in the credit scoring community are typically structured in a format featuring a mixture of numerical and categorical attributes \cite{wang2015survey}. Unstructured data such as raw text \cite{gunnarsson2021deep} has potential in the credit rating domain; however, it has not been extensively explored  \cite{shi2018deepclue}. Hence, we review studies where unstructured data had been used for prediction and classification. Lee et al. \cite{lee2014importance} analyzed the importance of text in the prediction of stock price movement by building a corpus that facilitated the forecasting tasks. They showed that the incorporation of text in the prediction task could improve results significantly. Goldberg et al. \cite{goldberg2016primer}  found that CNNs can be particularly beneficial when the useful information is sparse and dispersed in different places in the data, which is typically the case with textual data. The claim was further supported by Mai et al. \cite{mai2019deep}, who applied CNN to textual data with accompanying accounting data and reported better  prediction accuracy for corporate bankruptcy. 
Li et al. \cite{li2020multimodal} used an LSTM-based model for stock prediction to address the issues caused by interactions among different modes and heterogeneity of the data that featured temporal data and online news media. 
 
In the last decade,  NLP problems have seen significant improvement using deep learning methods, particularly those incorporating attention mechanisms and Transformer models  inspired  from biological cognitive systems, in which specific words in a sentence receive more attention than others \cite{vaswani2017attention}. Transformer-based models feature encoder–decoder LSTM models have been used for credit risk application \cite{dorfleitner2016description,chen2018role,netzer2019words}. Kriebel et al. \cite{kriebel2022credit} evaluated Transformer-based  models for text extraction over credit-relevant information achieving better performance than traditional machine learning models  In addition, the authors noted that the importance of textual data for credit default prediction was reinforced by the significant decrease in the accuracy  when the text  data was removed. Similarly, there have been further studies regarding  textual information  in financial applications. Mai et al. \cite{mai2019deep} and Matin et al. \cite{matin2019predicting} demonstrated that the use of textual disclosure and segments is promising for the prediction of financial distress and bankruptcy. However, using different datasets, Dorfleitner et al. \cite{dorfleitner2016description} and Chen et al. \cite{chen2018role} found no clear evidence that text characteristics could predict credit default in peer-to-peer lending; this  contrasts Matin et al. \cite{matin2019predicting}, who improved bankruptcy prediction by adding textual information related to the firm’s annual reports. Finally, Stevenson et al. \cite{stevenson2021value} applied the BERT model to predict loan default by small companies using textual loan assessments provided by lenders. They concluded that although the text data was useful for prediction when being assessed alone, the value of text data remained unclear because addition of text data to the already existing structured data produced no improvement in aggregate performance. In general, we find that most all the mentioned studies demonstrated that the addition of text data to deep learning models as an additional data channel can lead to better performance. 


\section{Methodology}
\label{sec:methodology}

\subsection{Data}

We use four sources of numerical and textual datasets, matched according to time index, and two identifiers, namely the company’s symbol and CUSIP (Committee on Uniform Securities Identification Procedures) identity (ID). The CUSIP ID features 9 digit combination of letters and numbers that serves as a unique identifier for securities such as stocks, bonds, and other financial products.
Regarding the data sources, for the structured data, we utilized the Wharton Research Data Services (WRDS), which is a private subscription-based data repository widely recognized for its comprehensive financial, marketing, and economic information. Specifically, we accessed the following databases within WRDS: 
\begin{itemize}
    \item  CRSP (Center for Research in Security Prices): provides market-related data including stock prices and returns.
    \item Compustat: provides statistics about corporate financial ratios and credit ratings essential for our financial analysis.
    \item Mergent and TRACE (Trade Reporting and Compliance Engine): provides detailed bond information, which provided insights into debt securities and their performance.
\end{itemize}

Due to the proprietary nature of the WRDS datasets, we cannot publicly share the raw data.  We provide a document that describes each data source comprehensively, outlines our data collection processes, and details the features and characteristics of the datasets on our GitHub \footnote{\url{https://github.com/Banking-Analytics-Lab/MultimodalFusionRatings}}.

The origin of unstructured data specifically refers to text documents from earnings call transcripts.  Additionally, to facilitate the use of our Python code for our framework, we created a synthetic dataset with similar structures and noise characteristics to the original. The small synthetic sample of the data that has been synthetically altered by large language models (Gemini) in our GitHub repo.
The dataset contains 27,854 valid records, and the input variables feature a minimum lag of two months compared to the target timestamp. We split the data randomly, using 20\% as the test set; the remaining data was split into 20\% as a validation set and 80\% as a training set.

\subsubsection{Target Data}
The importance of credit rating prediction lies in the crucial information it provides, enabling knowledgeable borrowing and investment decisions, as well as market regulation and risk management. Several credit rating agencies exist, but the three largest and best known (Fitch Ratings, Moody’s, and S\&P) control approximately 95\% of the rating market \cite{white2010markets}. We used a mapping table that presents different rating types had equivalence under the European Union Credit Rating Agency Regulation (CRAR) \cite{jewell1999comparison}. 

\begin{figure}[htbp!]
\centering
\includegraphics[width=3in]{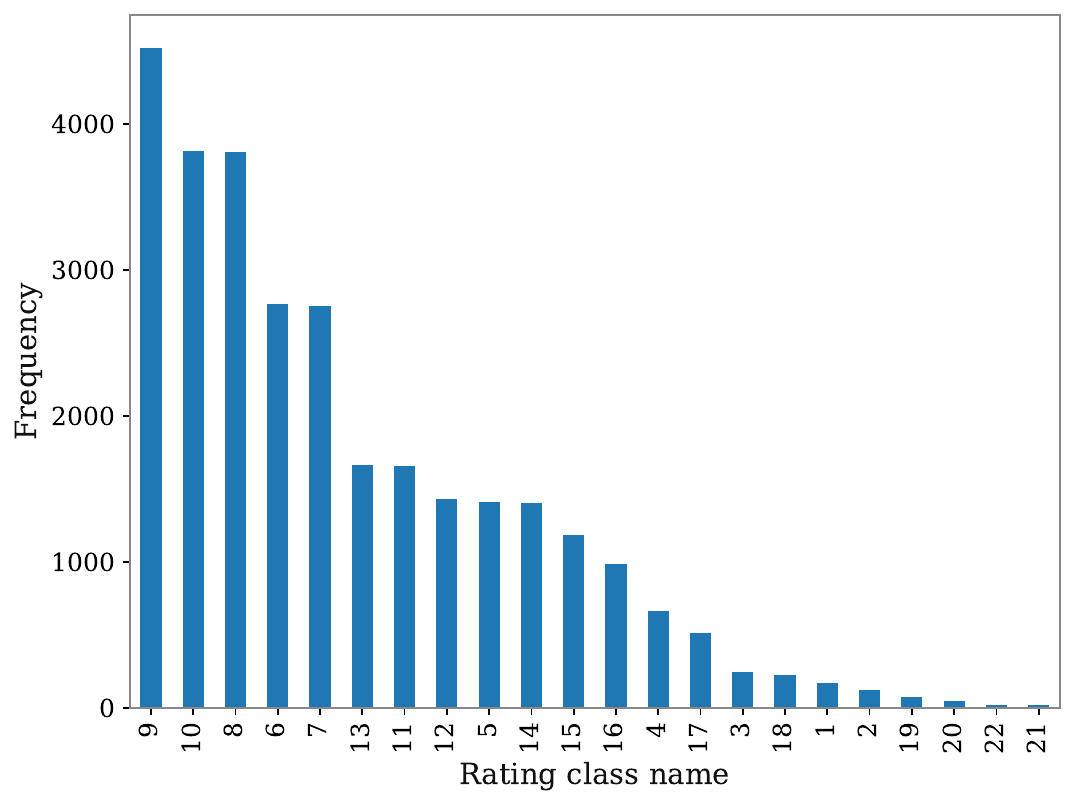}
\caption{Frequency of ratings in the original dataset, where some classes have a disproportionately low number of ratings, suggesting that they may not contain sufficient information to support accurate classification. }
\label{Fig.3}
\end{figure}

Table~\ref{tab:data} links letter-based credit ratings with numerical values to provide a means of mapping creditworthiness onto a quantitative scale. Typically, credit rating agencies use a letter-based system to denote credit quality, employing combinations of letters such as AAA, AA, A, BBB, BB, and so on. The highest credit quality is symbolized by the AAA rating in both Fitch and S\&P rating systems. By using the conversion table, users can accurately convert these letter-based ratings into a numerical scale, which is valuable for performing quantitative analysis or modelling. Based on the table, there are 22 rating classes, in which class 1 refers to the highest quality and class 22 refers to the defaulted companies. We merged the low-frequency classes (below 5\%) with their closest classes since the amount of data in some classes was significantly lower than in the other classes (Figure~\ref{Fig.3}), which helps to avoid class imbalance problems and increase statistical power.   Table~\ref{tab:data} indicates which classes were merged with each other.

\begin{table}[htbp!]
\small
	\begin{center}
		\caption{Conversion of letter-based credit ratings to numeric codes. The last two columns indicate the newly merged classes and their frequency  (source for first three columns: \cite{jewell1999comparison})}
		\label{tab:data}
		\adjustbox{max width=\textwidth}{
		\begin{tabular}{@{}ccccc@{}}
			\toprule
			 Code & Moody's & Fitch and S\&P&New class &Frequency \\\midrule\midrule
			 1& Aaa & AAA & \multirow{5}{*}{1} & \multirow{5}{*}{9\%}\\
			 2& Aa1 & AA+ & \\ 
			 3& Aa2 & AA & \\ 
			 4& Aa3 & AA-- & \\  
			 5& A1 & A+ & \\ \midrule
			 6& A2 & A &2 & 9\% \\ \midrule
			 7& A3 & A-- &3& 9\% \\ \midrule
			 8& Baa1 & BBB+ & 4& 13\%\\ \midrule
			 9 & Baa2 & BBB& 5& 15\% \\ \midrule
			 10 & Baa3 & BBB-- & \multirow{3}{*}{6}& \multirow{3}{*}{23\%}\\ 
			 11 & Ba1 & BB+ &  \\  
			 12 & Ba2 & BB & \\ \midrule
			 13 & Ba3 & BB-- & \multirow{3}{*}{7}& \multirow{3}{*}{14\%}\\ 
			 14 & B1 & B+ & \\ 
			 15 & B2 & B &  \\ \midrule
			 16& \multirow{2}{*}{B3} & B-- & \multirow{7}{*}{8}& \multirow{7}{*}{6\%}\\ 
			 17 & & CCC+&  \\ 
			 18 & Caa & CCC &  \\ 
			 19 &\multirow{2}{*}{Ca}& CCC-- &\\ 
			 20 && CC& \\ 
			 21 & C & C &\\
			 22 & D & D &  \\ \bottomrule
		\end{tabular}
	}
	\end{center}
\end{table}

\subsubsection{Numerical Channels}

In the case of the numerical channel data, they represent the following quantities: 
\begin{enumerate} 
   \item 	\emph{Bond historical} contains eight features related to the securities such as monthly high price, low price, trading volume, and other trading summaries.

   \item 	\emph{Financial ratios} includes 45 features related to the financial ratios of the companies including valuation, profitability, capitalization, liquidity, and efficiency ratios.

   \item  The \emph{market} dataset contained 98 features related to monthly indexes built on market capitalization and portfolios on S\&P index, stock market returns, and U.S. Treasury and inflation indicators.

   \item The \emph{covariate features} dataset contains the rating agency type and the last observed rating class at that timestamp. 
\end{enumerate} 

\subsubsection{Text Data}
We gathered text documents related to earnings call transcripts provided for each company by the Seeking Alpha website \cite{seeking_alpha}. Seeking Alpha specializes in financial news and analysis; it covers a wide range of publicly traded companies, stocks, and investments, with specific emphasis on those based in North America, particularly the US and Canada. One of its standout features is its platform for earnings call transcripts \cite{french-marcelin2020rise}. An earnings call transcript is a written account of a company’s financial conference call with analysts and investors. It summarizes the company’s financial results and provides information regarding its performance, operations, and prospects. During the call, management provides a prepared statement, followed by a question and answer session with analysts and investors. The aim of the call is to offer a more in-depth understanding of the company’s financial situation and operations, and the transcript serves as a valuable resource for investors and analysts to access for a precise and complete record of the information shared \cite{li2020corporate}.

\begin{figure}[!h]
\centering
\includegraphics[width=3.5in]{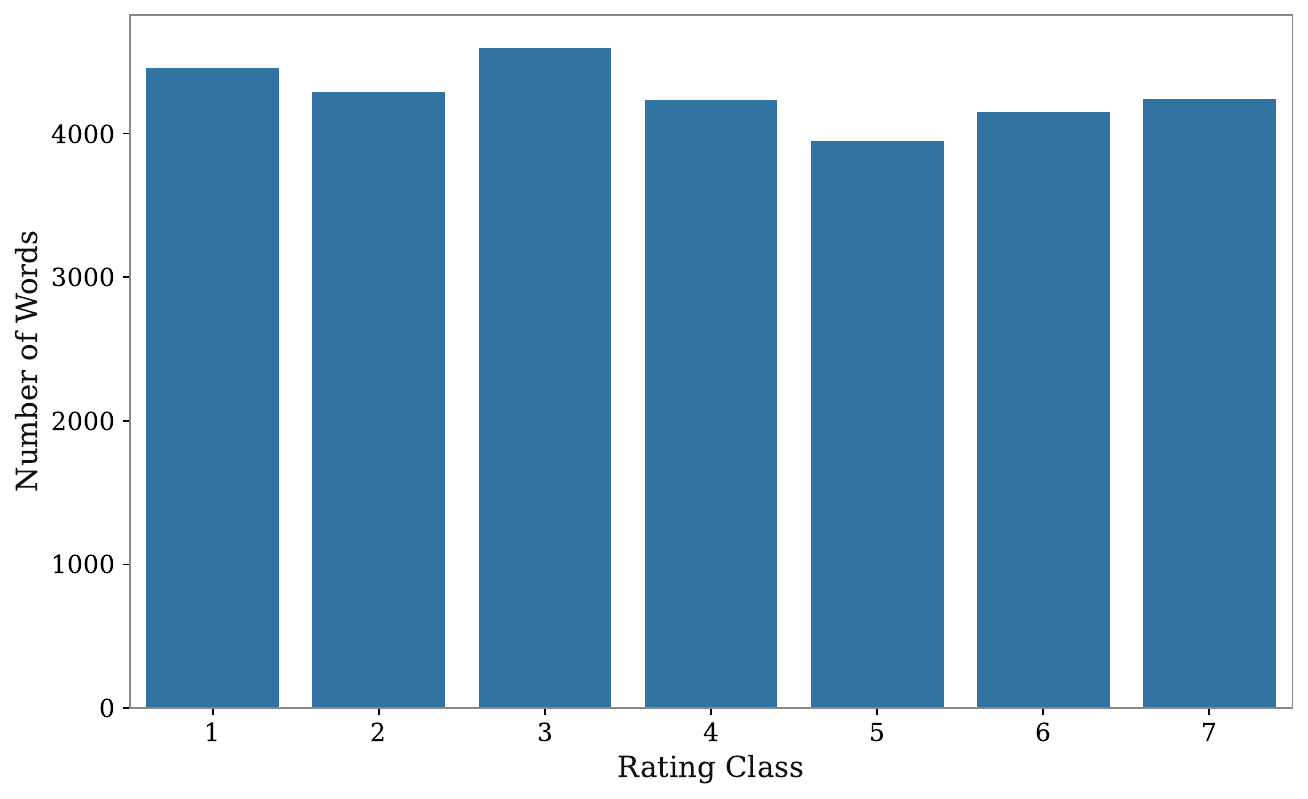}
\caption{Average number of words per earnings call transcript across different rating classes. }
\label{rating_freq}
\end{figure}

As shown in the Figure~\ref{rating_freq}, the average number of words per earnings call transcript is relatively consistent across different rating classes, with a slight variation. The number of words generally hovers around 4,000 across most classes, indicating that the amount of textual data is fairly balanced regardless of the company's credit rating. This consistency in word count suggests that the volume of information available does not vary significantly with the rating, allowing for a more uniform approach to processing and analysis across the dataset.
We used this text as an input channel for our models. To pre-process the text, we fixed various Unicode errors; applied lowercase; replaced all URLs, email addresses and phone numbers with a special token; and finally removed punctuation and stop words. The text was then tokenized into words using the Keras tokenizer \cite{kerasnlp2022} and the BERT encoder tokenizer \cite{bert_tokenizer}, which was used in conjunction with the BERT model. In BERT tokenization, text is split into smaller units called tokens, which consist of words and subwords. Each token is assigned a unique integer identifier for the model to recognize. This approach enhances the model’s understanding of each word’s meaning and the relationships between words in a sentence. The tokenization process also incorporates special characters, such as punctuation, and adds specific tokens to mark the start and end of a sentence \cite{lu2022comparative}. 

We used the Keras embedding tokenizer for the other models (CNN, LSTM, and GRU). The Keras tokenizer performs tokenization by transforming words or subwords into integer tokens. It initially converts all text to lowercase, then splits the text into separate words. Each word is then assigned a unique integer based on its frequency in the dataset. Once the text is tokenized, it can be used as input to a deep learning model for further analysis in NLP applications \cite{chollet2021deep}.

\subsection{Pre-trained language models}

Transformer-based models  \cite{vaswani2017attention} have been commonly used in NLP tasks such as machine translation \cite{nguyen2020neural} and question-answering \cite{namazifar2021language}. The transformer model is based on an encoder–decoder LSTM network with an attention mechanism that significantly increases the number of trainable parameters compared  to  conventional LSTM models, and also provides better results, particularly for tasks that are complex and require large scale datasets.
The BERT model \cite{devlin2018bert} is a transformer-based model that learns bidirectional representations from both the left and the right context of a word within a sentence to overcome the limitation of unidirectional language models. The   BERT model has been pre-trained on a large corpus that includes the Wikipedia English dataset along with the BookCorpus. Due to its state-of-the-art performance in NLP tasks such as corpus classification \cite{devlin2018bert} and text classification \cite{sun2019fine}, we use the BERT-base, with all its trainable parameters frozen, as one of the models for processing the textual channel of the credit rating data in our proposed framework. In this study, we employ an attentive CNN, which combines a CNN model for numerical channels with an added attention module in its architecture. This approach can improve the prediction result in classification applications \cite{neumann2017attentive}.

\subsection{Information Fusion}

In addition to applying deep learning models to structured and unstructured data, we further investigate the impact of different information fusion strategies in our proposed framework. 
We implement two stages of fusion strategies, including   \textit{fusion level} and \textit{fusion techniques}.
 
\subsubsection{Fusion Level} 

\begin{figure*}[htbp!]
\centering
{\includegraphics[width=5in]{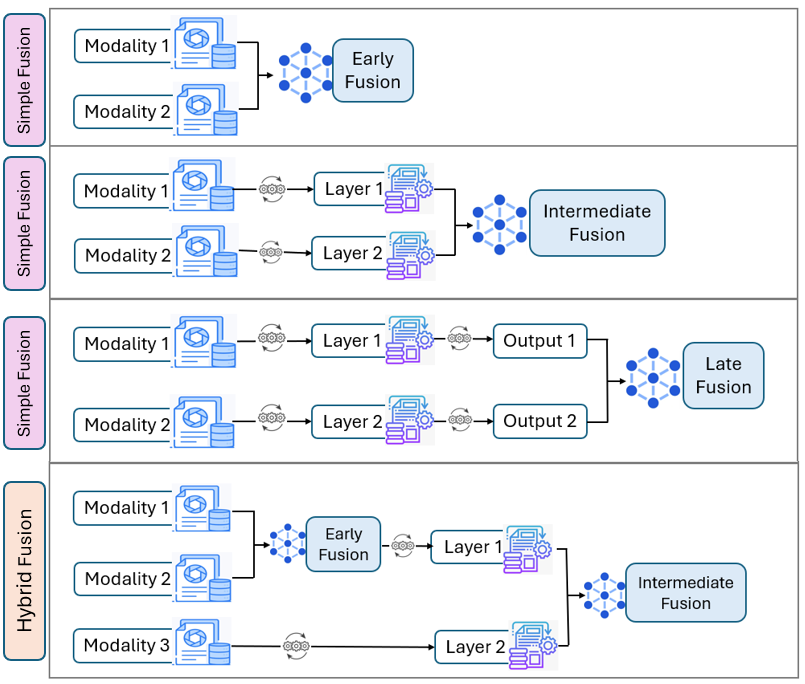}
\caption{This diagram illustrates various fusion strategies at different levels: Early, Intermediate, and Late Fusion. The Simple Fusion strategy is depicted in the first three sections, where only two modalities are involved, and the fusion occurs at a single level—either early, intermediate, or late. In the Hybrid Fusion strategy (last section), multiple modalities are combined at multiple levels (both early and intermediate), leveraging the strengths of each approach.}
\label{Fig.1}}
\end{figure*}

In our framework, fusion refers to the strategies used to combine information from multiple modalities, such as concatenation of information from multiple streams of data. The fusion level refers to the stage in the pipeline at which the information is merged, which can occur at various points, from the initial raw data input to the final stages of prediction. We present the different fusion levels below:
\begin{itemize}
    \item \emph{Early Fusion} involves merging information from multiple modalities at the very beginning of the processing pipeline after  data extraction. Early fusion allows us to capture diverse and unaltered information directly from the source. In Figure~\ref{Fig.1}, this is depicted as combining Modality 1 and Modality 2 and feeding the combined data into a model.  
    \item \emph{Intermediate Fusion} involves the combination of data at an intermediate stage, typically after some level of feature extraction has already occurred. This allows us to work with higher-level features that have already undergone some degree of processing and refinement. As shown in Figure \ref{Fig.1}, intermediate fusion merges the outputs from layers of neural networks processing Modality 1 and Modality 2 separately before further processing.
    
    \item \emph{Late Fusion} delays the combination of information until the final stages of processing. Typically, late fusion involves merging the predictions or decisions from separate models that have processed different modalities independently. In Figure \ref{Fig.1}, the outputs (Output 1 and Output 2) from separate models based on Modality 1 and Modality 2 are merged to form the final prediction.
    
    \item \emph{Hybrid Fusion} combines multiple fusion strategies as shown in Figure \ref{Fig.1}, where  Modality 1 and Modality 2 are combined (Early Fusion), while Modality 3 is processed separately and  then combined (Intermediate Fusion).
\end{itemize}

\subsubsection{Fusion in Terms of Techniques} 

\begin{figure*}[htbp!]
\centering
{\includegraphics[width=4in]{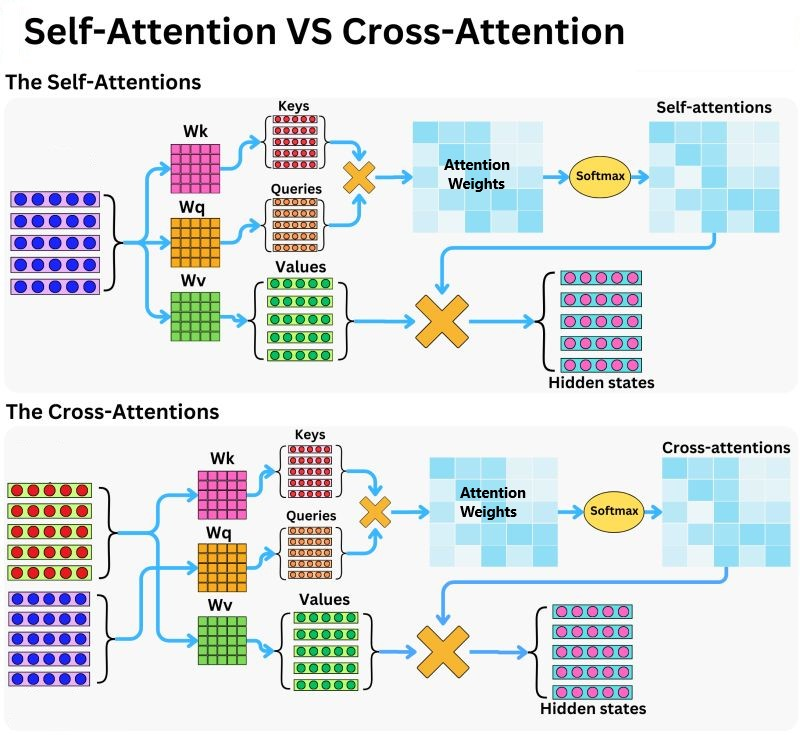}}
\caption{The diagram illustrates Self-Attention (top), where elements of a sequence attend to each other, and Cross-Attention (bottom), where one sequence (query) attends to another sequence (keys and values). The three vectors are used for data processing include query (Q), key (K), and value (V). Self-Attention captures internal dependencies within a single sequence, while Cross-Attention links information across different sequences, crucial for tasks like multi-modal learning and sequence-to-sequence models.} 
\label{Fig.tech}
\end{figure*}

In the literature, several  fusion techniques are prominent \cite{cheng2024multi, tan2024asset, che2024predicting} in combining information from multiple modalities effectively. These techniques include simple approaches such as  concatenation and more complex mechanisms such attention-based fusion, discussed below: 
\begin{itemize}
    \item \textit{Concatenation} combines information coming from different modalities of data to form a single feature vector. The data modalities can be of the same nature or different nature, i.e. text data can be concatenated with numeric data and processed further. 
    
    \item \textit{Cross-Attention Fusion} is an advanced technique inspired by the self-attention mechanism commonly used in deep learning models such as transformers \cite{shaw2018self,vaswani2017attention}. In self-attention, three vectors are used for data processing: the query ($Q$), key ($K$), and value ($V$) (Figure \ref{Fig.tech}). In cross-attention, the process is similar, but with a key difference: the query and value vectors are generated from one modality, while the key vector comes from another modality. This allows the model to capture the interactions and correlations between different modalities more effectively than simple concatenation.
\end{itemize}

In our multimodal framework, we consider several modalities of data, including text and numerical data that are utilized simultaneously. We use three fusion levels (early, intermediate, hybrid) and two fusion techniques (concatenation, cross-attention) to determine which combination of strategies can better capture the information for the prediction of ratings. We identify the optimal approach for integrating diverse data sources to enhance the accuracy and robustness of our  model by evaluating the combination of these different fusion strategies.
 
\subsection{Multimodal deep learning framework}
 
We revisit our goals to highlight that our study makes several contributions to the field of credit rating prediction using deep learning with multimodal data. First, we systematically explore and compare various multimodal fusion strategies (such as early/hybrid fusion and cross-attention fusion) to understand their impact on model performance. Our framework includes a novel exploration of structural modes within different deep learning models to ascertain the optimal fusion strategy. We also quantify the importance of each data channel within our models, providing insights into how structured and unstructured data can be effectively integrated. 
Lastly, we analyze the impact of the COVID-19 pandemic on our model predictive performance to review  adaptability to external shocks and their capacity to leverage significant external factors for improved prediction accuracy.

Our framework features appropriate data structures and techniques to address which combination of fusion strategies and deep learning models could capture more useful information from the data for model prediction. Hence, we investigate three areas, namely 1) deep learning models, 2) fusion-type strategies including concatenation and cross-attention fusion, and 3) levels of fusion involving early, intermediate, late and hybrid fusion. 

Our study  deals with both structured and unstructured data, necessitating an architecture that can effectively harness the diverse information these data types offer.  Our literature review revealed that multimodal models are frequently employed in financial applications (e.g. \cite{zhang2024smpdf}), prompting us to explore how their combination might enhance information capture for credit rating purposes. Therefore, we will conduct experiments to determine which model combinations, the method of combination (either simple concatenation or cross-attention fusion), and the stage of integration (early or intermediate) would yield the optimal results for our study's objectives.

\begin{figure*}[!h]
\centering
\subfloat[Group 1: Simple Concatenation]{\includegraphics[width=2.2in]{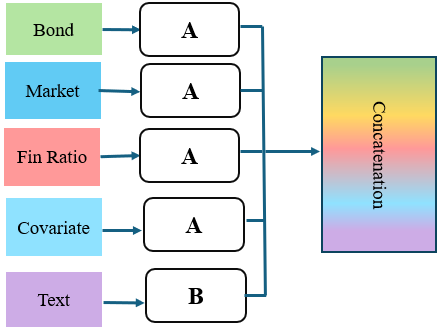} 

\label{Fig.2a}}
\hfil
\subfloat[Group 2: Simple Concatenation-Attention]{\includegraphics[width=2.3in]{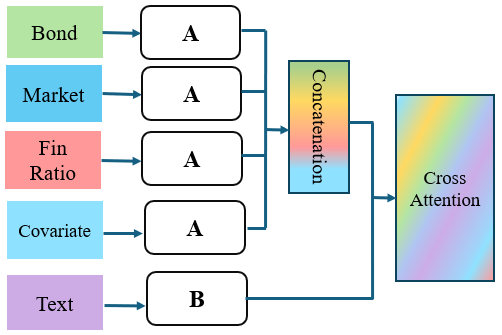}
\label{Fig.2b}}
\\
\subfloat[Group 3: Hybrid Concatenation]{\includegraphics[width=2.2in]{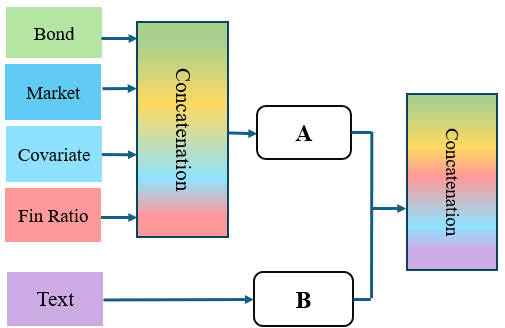}
\label{Fig.2c}}
\hfil
\subfloat[Group 4: Hybrid Concatenation-Attention]{\includegraphics[width=2.2in]{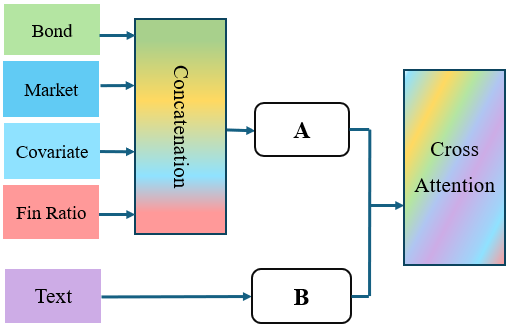}
\label{Fig.2d}}

\caption{The four multimodal frameworks for credit rating prediction with different combinations of deep learning-based submodels (A and B)  with fusion type and fusion level strategies.  We present the  submodel  architectures and implementation in Table~\ref{tab:combination}.}
\label{fig:framework}
\end{figure*}

 We  employ  CNN, Convolutional LSTM (ConvLSTM), Convolutional GRU (ConvGRU), CNN with Attention mechanism (CNN-Attn) and BERT models in our framework for credit rating prediction as shown in Figure~\ref{fig:framework}. 
 The five channels of inputs include three semistructured channels (bond, market, financial ratios), one structured channel (covariate, which was previous rating information) and one text channel in each category. In each group of our multimodal framework, we combine different fusion levels (Figure~\ref{Fig.1}) and fusion techniques  (Figure~\ref{Fig.tech}) with selected deep learning models. Submodel A in Figure~\ref{fig:framework} represents the sub-model for numerical channels, and submodel B refers to a sub-model for the text channel. Each group (e.g. Group 1) features submodel A and B which can use different deep learning models, from the list that includes CNN, ConvGRU, ConvLSTM, CNN-Attn and BERT. We present the exact combination for each group, with model implementation details, in Table~\ref{tab:combination}.
 



After the fusion of outputs from submodel A (processing numerical data) and submodel B (processing text data), the combined output is passed through a series of layers to reach the final model output. Specifically, this fused output is first flattened, then processed through two dense layers (with 64 neurons in the first layer), followed by a dropout layer (with a dropout rate of 0.5) to prevent overfitting. 
We determined the configuration of these parameters by combining insights from literature review with extensive experimentation. Guided by existing research on deep learning architectures, we explored various configurations and systematically assessed their performance.
The final output layer is responsible for producing the model's predictions. These layers, consistent across all model groups (Group 1 to Group 4), have been omitted from Figure~\ref{fig:framework} for simplicity.


Next, we present the fusion characteristics of each group in Figure~\ref{fig:framework} highlighting  submodel  architectures and fusion levels:

\begin{itemize}
\item Simple Concatenation (\textbf{Group 1}): we pass each numerical dataset (channel)  through submodel A (e.g.  independent CNN models), and use submodel B for the text data stream. We then concatenate their penultimate layers using simple intermediate-level fusion, as shown in Figure 1. 

\item  Simple Concatenation-Attention (\textbf{Group 2}): The initial phase involves training four distinct numeric channels individually through submodel A, while the text channel undergoes training via sub-model B. After model training, we concatenate the penultimate layer of the numeric modalities. In the final step, we execute the fusion  between the text layer and the concatenated layer employing the cross-attention technique. Therefore, this architecture utilizes a simple Concatenation-Attention fusion approach. 

\item  Hybrid Concatenation (\textbf{Group 3}): This architecture employs a hybrid fusion approach where all numerical datasets undergo early fusion, functioning as inputs for sub-model A. Conversely, text data undergoes training via submodel B. Subsequently, we execute intermediate fusion   and the penultimate (next to last) layers of distinct modalities by concatenation of their corresponding output layers as shown in Figure 1. 
 
\item Hybrid Concatenation-Attention (\textbf{Group 4}): Following a pattern similar to Group 1, we implement a hybrid architecture in Figure \ref{fig:framework} with an initial fusion of all numerical datasets  (signal fusion), resulting in the concatenation of the numerical channel for  training of sub-model A. Meanwhile, the text channel was directed to submodel B, and the output layers were combined utilizing cross-attention fusion. 
   
\end{itemize}

We employ five deep learning-based submodels including CNN,  ConvLSTM,  ConvGRU,  CNN-Attn, and BERT models in our framework for credit rating prediction as shown in Figure~\ref{fig:framework}. As outlined in Table \ref{tab:combination}, we designed submodel A for processing numerical data, while submodel B is designated for handling text data (Figure \ref{fig:framework}).  
According to Table \ref{tab:combination}, there are four possible configurations for submodel A and submodel B with details given below.

\begin{itemize} 
\item \textbf{Combination I} features a CNN with two hidden layers for the numerical data streams in submodel A, and submodel B consists of a CNN with three hidden layers for the text data stream. Both submodels include dropout-based regularization (Drop) and max pooling (MaxP) layers between each layer, with global average pooling (GlobAve) applied at the final layer.

\item \textbf{Combination II}   employs ConvLSTM in submodel A by using a single convolutional layer with  an LSTM model for numerical data streams. Submodel B employs ConvLSTM with two convolutional layers with LSTM for the text data stream.

\item \textbf{Combination III} features ConvGRU in submodel A, which is similar to ConvLSTM, but with the LSTM  model replaced by a GRU. Submodel B features a ConvGRU with two convolutional layers for the text data stream.

\item \textbf{Combination IV} features CNN-Attn for submodel A that integrates an attention mechanism into a CNN model for the numerical data streams while  submodel B features the BERT model for the text data stream.
\end{itemize}

In our framework, text data undergoes several preprocessing steps before being input into the CNN-based models:

\begin{enumerate}

  \item  Text Preprocessing: We convert the raw text into lowercased and remove stopwords. We then  apply stemming, which reduces the inflected form of a word to one so-called “stem”, also known as a “lemma” in linguistics \cite{Singh2016}. These operations ensure consistency and reduce noise in the data.

  \item Tokenization: We use the tokenizer from Keras for tokenizing the text. Specifically, we utilized the Tokenizer class from the keras.preprocessing.text module\cite{keras_tokenizer} to convert the text into sequences of integer tokens, where each word is assigned a unique integer based on its frequency in the dataset. We then use padding to ensure uniform input length for the CNN model.
  
  \item  Word Embedding: We then pass the tokenized sequences   through an embedding layer, where each word is transformed into a dense vector representation. The embedding layer is trained as part of the model, learning meaningful word relationships.
\end{enumerate}

\begin{table*}[htbp!]
\small
\begin{center}
\caption{The architecture of submodels A and submodel B for each base model. Note that the implementation details feature the number of units in brackets for convolutional (Conv), max pooling (MaxP) and the Dropout (Drop), e.g. convolutional layer with 30 units is Conv(30). Similarly, we provide a number of hidden units LSTM (h), GRU(h)  and the number of heads and key dimensions in the attention layer (Attn(heads, key)).  The GlobAve operation is the final layer which combines the information from the previous layer (as shown in Figure \ref{fig:group3comb1} and \ref{fig:group3comb2})for the Concatenation operation, with implementation details shown. 
}

\label{tab:combination}
\begin{tabular}{>{\centering\arraybackslash}m{1.5cm} >{\centering\arraybackslash}m{2cm} >{\centering\arraybackslash}m{3.5cm} >{\centering\arraybackslash}m{2cm} >{\centering\arraybackslash}m{3cm}}
\toprule
Combination & Submodel A \newline Name & Submodel A \newline Implementation & Submodel B \newline Name & Submodel B \newline Implementation \\\midrule\midrule
I & CNN & Conv(64)$\rightarrow$ MaxP(2)$\rightarrow$ Conv(64)$\rightarrow$GlobAve & CNN & Conv(64)$\rightarrow$Drop(0.1)$\rightarrow$ Conv(64)$\rightarrow$MaxP(2)$\rightarrow $ Conv(64)$\rightarrow$GlobAve \\\midrule
II & ConvLSTM & Conv(64)$\rightarrow$ MaxP(2)$\rightarrow$\newline LSTM(32)$\rightarrow$GlobAve & ConvLSTM & Conv(64)$\rightarrow$Drop(0.1)$\rightarrow$ \newline Conv(32)$\rightarrow$Drop(0.1)$\rightarrow$ \newline Conv(16)$\rightarrow$MaxP(2)$\rightarrow$ \newline LSTM(32)$\rightarrow$GlobAve \\\midrule
III & ConvGRU & Conv(64)$\rightarrow$MaxP(2)$\rightarrow$\newline GRU(128)$\rightarrow$GlobAve & ConvGRU & Conv(64)$\rightarrow$Drop(0.1)$\rightarrow$\newline Conv(32)$\rightarrow$MaxP(2)$\rightarrow$\newline Conv(64)$\rightarrow$Drop(0.1)$\rightarrow$\newline GRU(128)$\rightarrow$GlobAve \\\midrule
IV & CNN-Attn & Conv(64)$\rightarrow$MaxP(2)$\rightarrow$\newline Conv(64)$\rightarrow$MaxP(2)$\rightarrow$\newline Attn(1,5)$\rightarrow$GlobAve & BERT & BERT \\\bottomrule
\end{tabular}
\end{center}
\end{table*}

\subsection{Evaluation metrics}
We select a set of criteria to assess the performance and the statistical prediction power of each model. The Receiver operating characteristic (ROC) curve is a graph showing the performance of a classification model at all classification thresholds. The AUC represents the Area under the ROC curve. The AUC is a widely used evaluation metric for binary and multi-class classification problems. It measures the ability of the model to distinguish between classes. The AUC is particularly useful because it is scale-invariant and classification-threshold-invariant\cite{davis2006relationship}, meaning it evaluates the model's performance across all classification thresholds.

The F1 score is a weighted score of Precision and Recall, which are widely used for class-imbalanced problems.  The F1 score is the harmonic mean of precision and recall, providing a single metric that balances both false positives and false negatives. It is particularly useful when the class distribution is imbalanced, as is often the case in credit rating datasets, where certain rating classes may be underrepresented and  calculated as follows:

\begin{equation}
F-1=\frac{2}{recall^{-1}+precision^{-1}}
=\frac{2 \mathrm{TP}}{2 \mathrm{TP}+\mathrm{FP}+\mathrm{FN}}
\end{equation}
where TP is the number of true positives, FP is the number of false positives, and FN is the number of false negatives.

The Area Under the Precision-Recall curve (AUP-RC) \cite{davis2006relationship}  provides precision and recall across different thresholds, which is useful for class-imbalanced datasets, as it provides additional performance information on the minority class. Since our problem involves multi-class classification, we use a weighted average of AUC for each class versus the rest of the classes (one-versus-all approach) \cite{{hand2001simple}}.

\subsection{Experiment Setup}
\begin{figure*}[htbp!]
\centering
\includegraphics[width=4.4in]{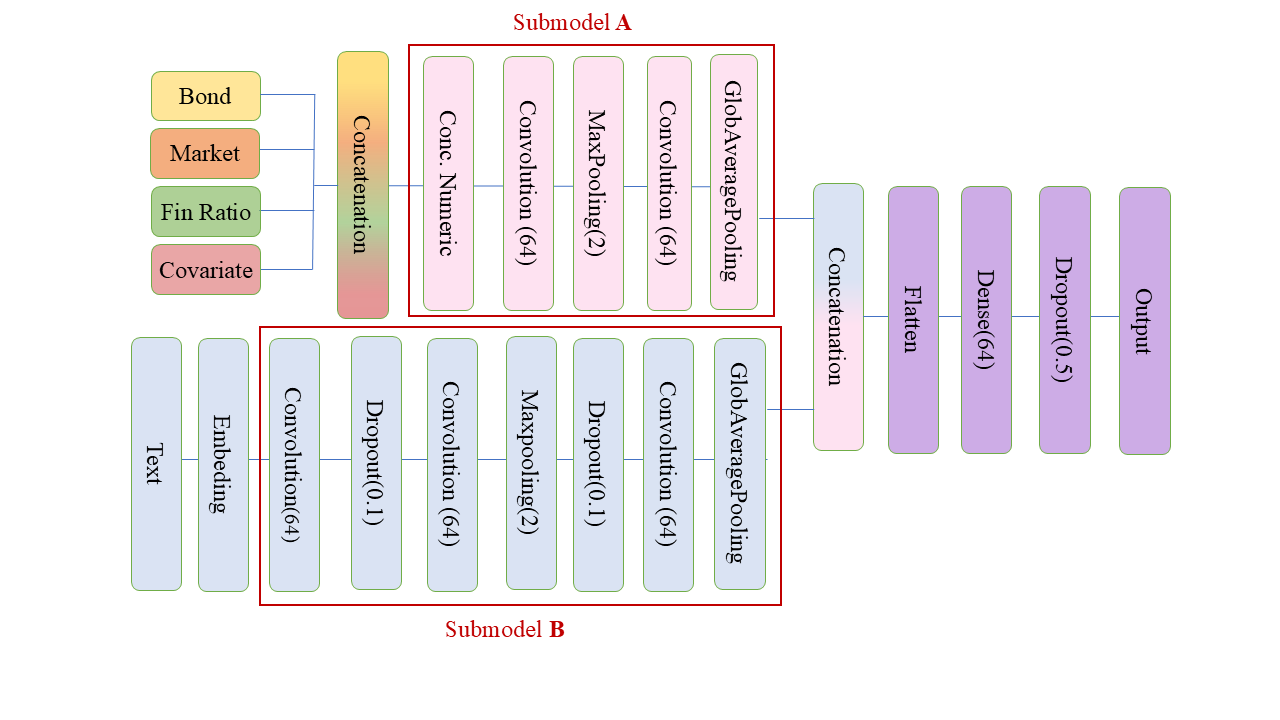}
\caption{Hybrid Concatenation (Group 3 in Figure \ref{fig:framework}) featuring  Combination I (Table \ref{tab:combination}) showing the convolution and dropout layers with two streams of data that include text and numerical data.} 
\label{fig:group3comb1}
\end{figure*}

\begin{figure*}[htbp!]
\centering
\includegraphics[width=4.4in]{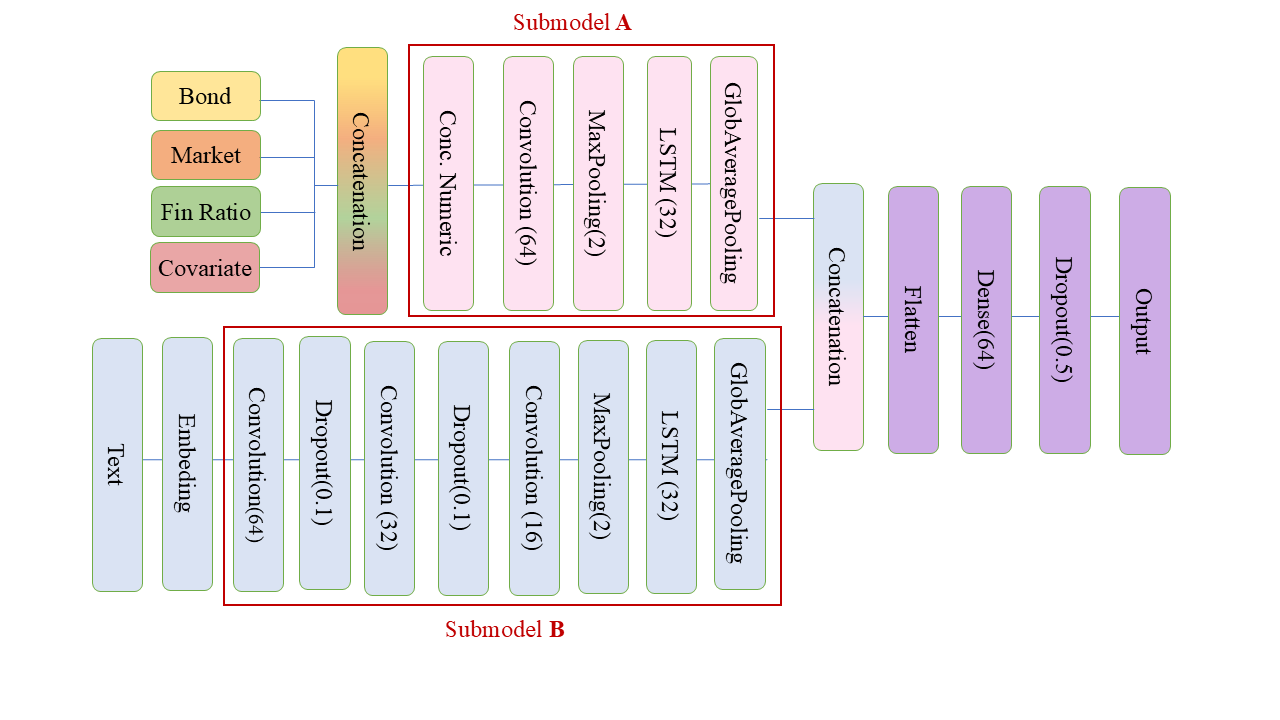}
\caption{Hybrid Concatenation (Group 3 in Figure \ref{fig:framework}) featuring  Combination II (Table \ref{tab:combination}) showing  the convolution and dropout layers with the LSTM model for two streams of data that include text and numerical data.} 
\label{fig:group3comb2}
\end{figure*}

We use  16 distinct models in our multimodal deep learning framework, comprising the four combinations of submodels in Table~\ref{tab:combination} and the four groups of submodels with fusion strategies given in Figure~\ref{fig:framework}. We conducted extensive hyperparameter optimization to ensure optimal performance and presented the details for each model architecture in Table~\ref{tab:combination}. The implementation details feature the number of units in brackets for convolutional (Conv), max pooling (MaxP) and Dropout (Drop), e.g. convolutional layer with 64 units is Conv(64). Similarly, we provide a number of hidden units in LSTM(h), GRU(h)  and the number of heads in the attention layer and the dimensionality of each key vector (Attn(heads, key)). The GlobAve operation is the final layer which combines the information from the previous layer (as shown in Figure \ref{fig:group3comb1} and \ref{fig:group3comb2}) for the Concatenation operation, with implementation details shown(Table~\ref{tab:combination}, Figures~\ref{fig:group3comb1},\ref{fig:group3comb2}).

We present the model architectures and explanations in the folder of Model's architecture in our GitHub repository.
(\href{https://github.com/Banking-Analytics-Lab/MultimodalFusionRatings}{Model's Architecture}).
This section outlines the general approach to hyperparameter selection across all models.

In trial runs and with guidance from similar models in the literature, we determined the hyperparameters for different submodels including CNN, ConvLSTM, ConvGRU, and Attn-CNN as shown in Table~\ref{tab:combination}. We selected the hyperparameters based on a balance between training speed, performance accuracy and convergence stability. We employed a systematic approach using a hold-out validation set to assess the performance of each parameter configuration for the respective submodels. Hence, we also determined the convolutional layer filter size and kernel size for respective models (\ref{tab:combination}), activation function (ReLU), batch size (100), Adam optimizer \cite{kingma2014adam} learning rate (0.0001), and dropout rate (0.5) for regularization for all the respective submodels in Table \ref{tab:combination}.

Figure~\ref{fig:group3comb1} presents the Hybrid Concatenation (Group 3) in Figure \ref{fig:framework}) featuring  Combination I (Table \ref{tab:combination}) showing the convolution and dropout layers with two streams of data that include text and numerical data. We concatenate the predictions from both CNNs for the feedforward neural submodel with fully connected (dense) layers and the dropout regularization layer for computing the output. Similarly, Figure \ref{fig:group3comb2} presents Hybrid Concatenation (Group 3) featuring  Combination II, showing the convolution and dropout layers with the LSTM model for two streams of data.


In each model experiment, we conduct 30 independent training runs with different initial weights and biases, and report the mean and standard deviation of their performance on the test dataset.
We implement our framework in Python and use T4 GPUs (NVIDIA T4) with 16 GB memory, and a Linux operating system (Ubuntu 20.04). We provide open source code and synthetic data in the GitHub repository, with details at the end of the paper. 

\section{Results}
\label{sec:prediction}

\subsection{Text Visualization}

\begin{figure}[htbp!]
\centering
\includegraphics[width=3.3in]{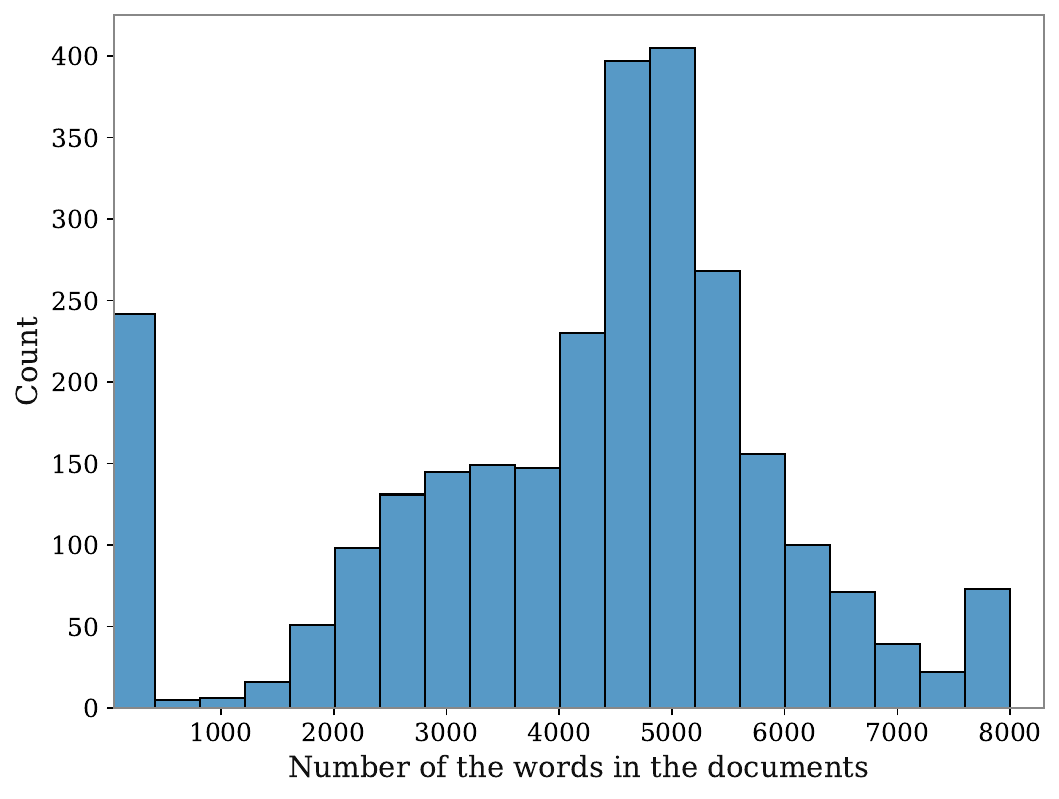}
\vspace{-0.3cm}
\caption{Frequency of the words in the documents: This histogram illustrates the distribution of document lengths within a dataset, where the X-axis represents the number of words per document, grouped into specific ranges (bins), and the Y-axis shows the count of documents that fall into each bin. }
\label{fig:datavisual}
\end{figure}

The text data for credit rating features documents that review company profiles and provides statements that can be helpful in understanding the relationship of earning calls with risk. The text dataset features 27,854 documents where the frequency of the words varies from 50 to 20,000 words as shown in Figure~\ref{fig:datavisual}  for each rating class on average. Note that as shown in Table~\ref{tab:data}, we initially had 22 classes, which were grouped to get 8 classes. 

An $n$-gram is a sequence of $n$ consecutive elements taken from a given sample of text   \cite{poliak2017efficient} and prominently used to visualize and analyze text datasets. We present  $n$-grams (bigrams and trigrams) to visualize the difference between the words in the high-rank (class ratings of 1 to 10 in Table~\ref{tab:data}) and low-rank classes  (class ratings of 11 to 22). 
High ratings, such as Moody's "Aaa" that corresponds to S\&P's "AAA" (listed under New Class 1 in the Table~\ref{tab:data}), represent the highest credit quality with the lowest risk. In contrast, low ratings are illustrated by entries like Moody's "C," equivalent to S\&P's "C" (categorized under New Class 8), indicating very high credit risk and representing the lowest tier of credit quality.

We split the documents by rating class for visualizing common words in the documents that were processed as had eight rating classes, as shown in Table \ref{tab:data}. However, we provide visualizations for two categories rather than eight, where Category 1 features the high ratings and Category 2 features the low ratings, to simplify the visualization.  
This implies that the documents in the dataset are divided  with ratings that indicate lower than median risk (Category 1), and   higher than median risk (Category 2).

\begin{figure*}[htbp!]
\centering
\subfloat[Bigram presenting the data for  the high-rank classes (Category 1)]{\includegraphics[width=2.6in]{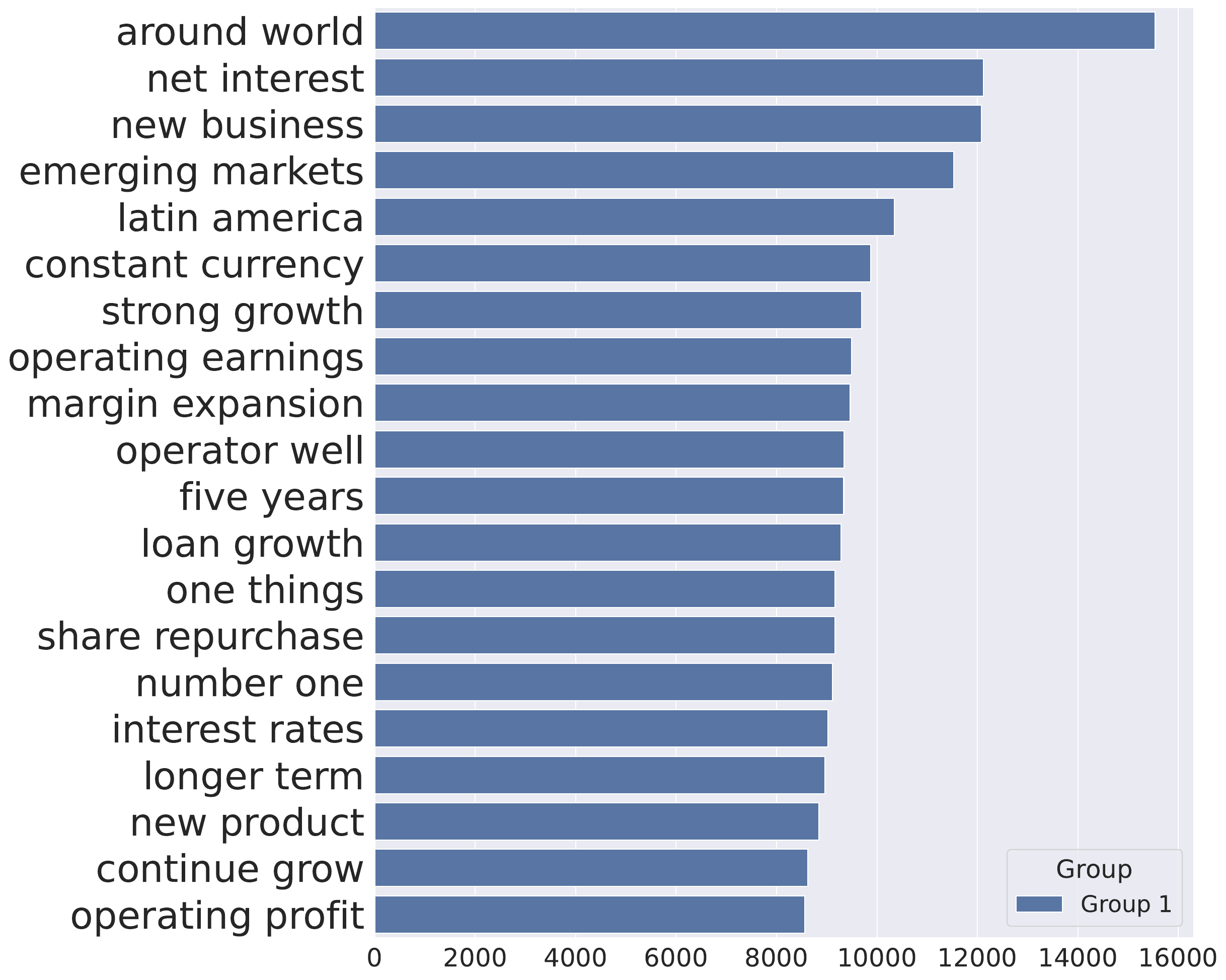} }
\hfil
\subfloat[Bigram  presenting the data for  the low-rank classes (Category 2)]{\includegraphics[width=2.6in]{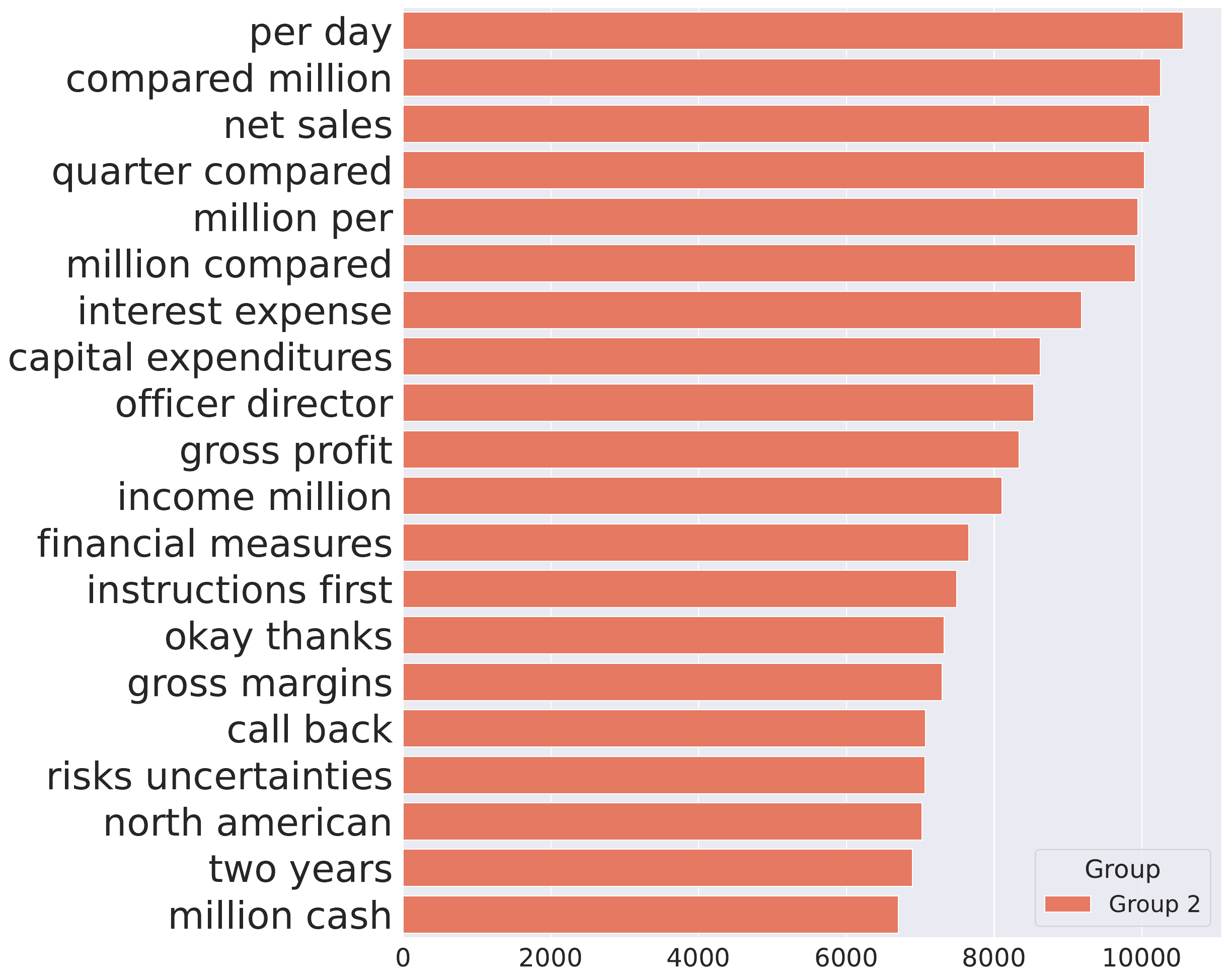} }
\caption{Bigrams for  high-rank classes (Category 1) and low rank classes (Category 2).}
\label{Fig.bi}
\end{figure*}

\begin{figure*}[htbp!]
\centering
\subfloat[Trigram featuring the high-rank classes]{\includegraphics[width=2.5in]{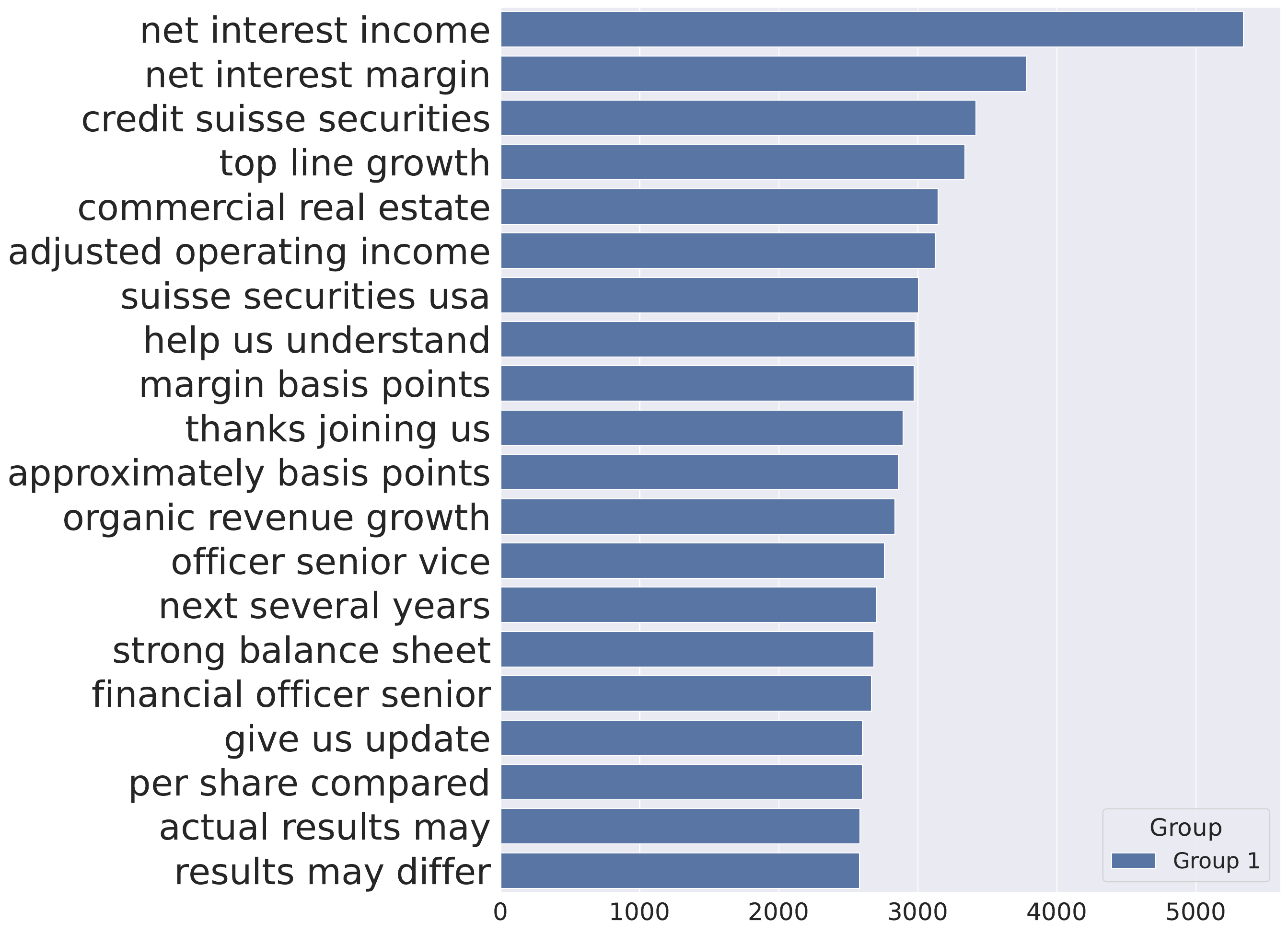}
\label{Group1}}
\hfil
\subfloat[Trigram featuring the low-rank classes]{\includegraphics[width=2.5in]{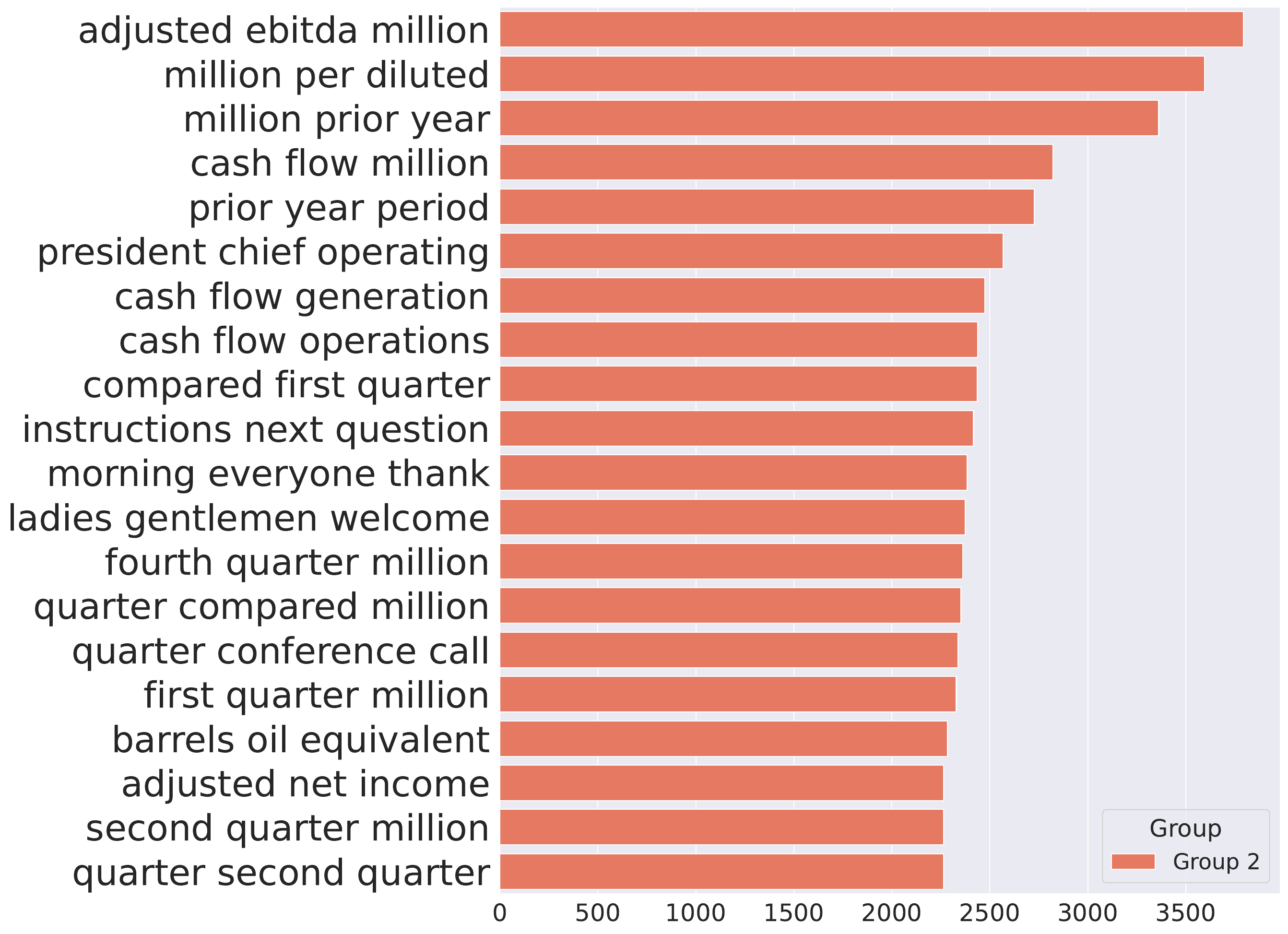}
\label{Group2}}
\caption{Trigrams for high-rank classes (Category 1) and low-rank classes (Category 2).}
\label{Fig.tri}
\end{figure*}

 
In the analysis involving both bigrams and trigrams (Figures~\ref{Fig.bi} and \ref{Fig.tri}), only those combinations of words that are unique to either the high or low credit rating groups are considered, excluding those that appear in both groups. This improves the distinction between the two categories by focusing solely on unique two-word (bigrams) and three-word (trigrams) sequences. The approach enhances the visibility of distinct linguistic patterns and phrase usage characteristic of each group, providing better insights into how language varies with credit risk levels.

In the high rating group (Category 1), phrases such as "around world," "net interest," and "new business" are prevalent, indicating a focus on global operations and financial growth. Conversely, in the low rating group (Category 2), unique phrases include "compared million," "net sales," and "gross profit," pointing towards a more localized or operational focus. 
Continuing our document analysis categorized by rating class, we present trigram analysis for deeper linguistic insights, as shown in Figure~\ref{Fig.tri}.

In the high rating group (Category 1), unique trigrams such as "net interest income," "adjusted operating income," and "strong balance sheet" suggest a financial narrative focused on growth, stability, and positive financial metrics. These phrases underline strategic financial management and robust economic health in organizations within this group.
Conversely, the low rating group (Category 2) predominantly uses trigrams like "adjusted ebitda million," "cash flow million," and "quarter compared million," which emphasize operational and performance-based metrics. These trigrams reflect a focus on operational outcomes and shorter-term financial performance, rather than long-term strategic positioning. 

\subsection{Model Prediction}

We present the results for the Groups (1, 2, 3, 4 in Figure \ref{fig:framework}) and Submodel Combinations (I, II, III, IV in Table \ref{tab:combination} to evaluate what type of deep learning models in numerical and text channels suit the data fusion strategies.  Table~\ref{tab:resultscomb} presents the performances of the respective models and their groups in terms of weighted-AUC, F1 score, AUP-RC, and computational time. We report  the mean and standard deviation based on 30 independent experiments for each model.  
As shown in Table~\ref{tab:combination}, the results reveal significant insights into the performance of various model configurations and fusion strategies. Note that, as shown in Table \ref{tab:data}, we have a multi-class problem with eight distinct classes. 

We find that the Hybrid Concatenation (Group 3 with Combination I) strategy in the framework provides the best performance. The hybrid concatenation features an architecture that combines hybrid early-intermediate concatenation. It uses submodel combination I featuring CNNs in submodel A and B, that incorporate four convolutional layers accompanied by dropout and max-pooling, along with global average pooling. 
In the case of Groups 1 and 2,  Combination I (CNN, CNN) and Combination  III (ConvGRU, ConvGRU)  produced the best performances in comparison to Combinations II and IV. In the Hybrid Concatenation-Attention Group (Group 4),  the performances of the respective submodels combinations (I-IV) are close to each other.
        

We note that Group 2 and 4 utilize cross-attention strategy, and we find this to be better than simple concatenation for Combination V (CNN-Atten, BERT). The maximum performance among all attention-based submodels (Combination V) occurred when each channel was separately trained and cross-attention was then applied to mix the outputs (Group 2).

Although both LSTM and GRU are based on RNNs, the performance of ConvGRU (Combination III) was much better than LSTM in all groups. One possible reason for this effect is that GRUs are simpler models, having only two gates, as opposed to LSTM networks. This simplicity may allow GRUs to more effectively capture dependencies in a dataset that is not excessively large, such as 27,000 records in our dataset.


The most effective strategy featured Combination I (CNN, CNN) with three hidden layers for the text channel and two for the numeric channel, with dropout layers and a dense neural submodel in the output layer as shown in Figure \ref{fig:group3comb1}.
 
\begin{figure}[htbp!]
\centering
\includegraphics[width=3.6in]{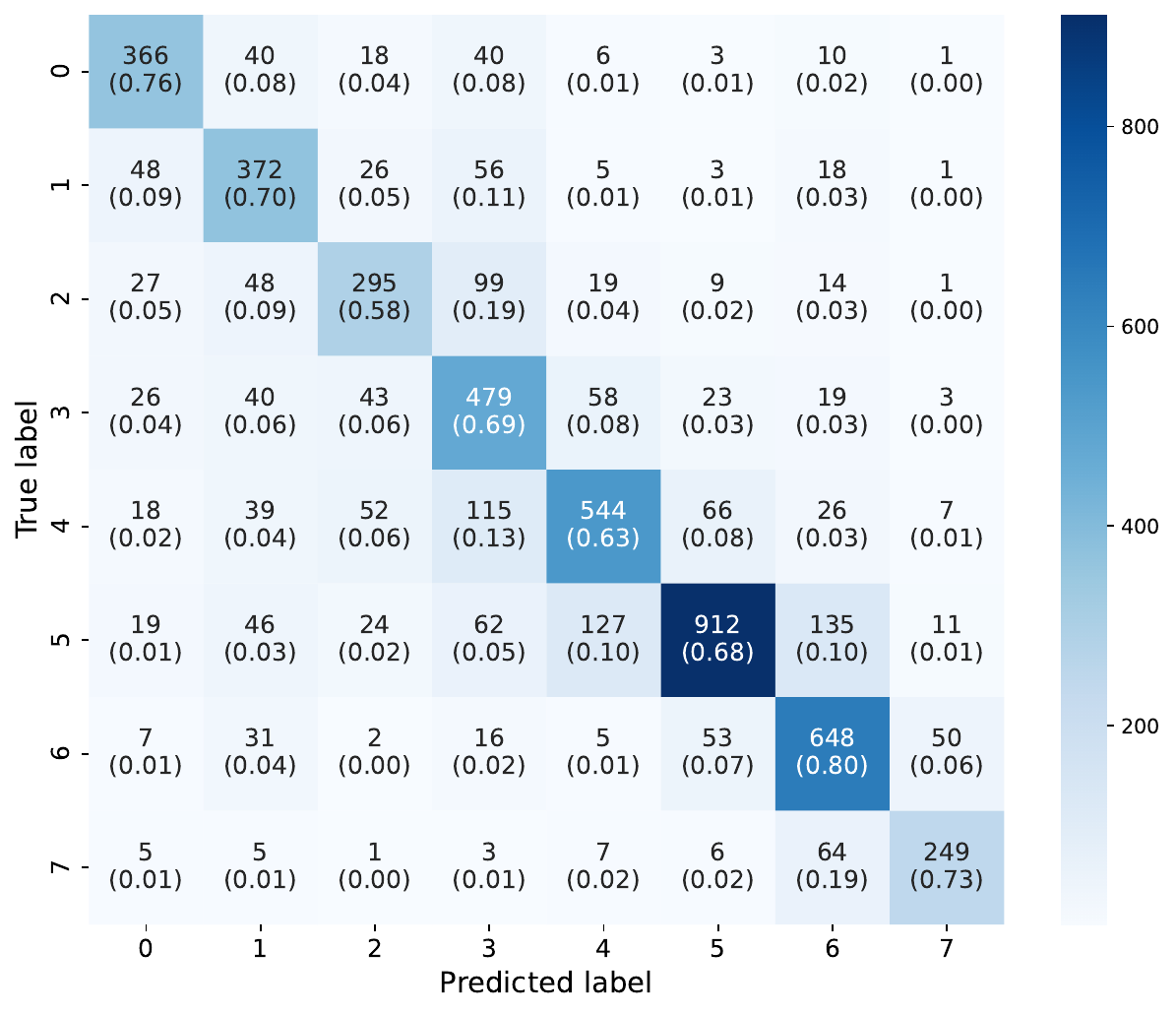}
\caption{Confusion matrix for the best model as shown in Figure \ref{fig:group3comb1}.}
\label{fig:conmat}
\end{figure}

The confusion matrix in Figure~\ref{fig:conmat} reports the best model from results earlier (Table \ref{tab:resultscomb}). The confusion matrix shows that the most accurate prediction belonged to the two highest and lowest quality bonds, specifically 80\% and 73\% for the two lowest quality classes and 76\% and 70\% for the highest rating, i.e. “AAA” to “A+” based on Fitch and S\&P agency ratings in Table~\ref{tab:data}). This indicates that the best model can accurately predict credit ratings with low and high credit quality, which are the most critical classes for investors.

\begin{table*}[htbp!]
\begin{center}
\caption{Model performance across different groups (Figure \ref{fig:framework} and  Submodel Combinations (Table \ref{tab:combination}), reporting Weighted-AUC, F1, and AUP-RC test data scores for 30 independent modal training experiments. Note that we only provide the mean for the elapsed time, given in minutes. }
\label{tab:resultscomb}
\adjustbox{max width=\textwidth}{%
\begin{tabular}{>{\centering}m{0.5cm} >{\centering}m{2.8cm} >{\centering}m{2.8cm} >{\centering}m{2.8cm} >{\centering}m{2.8cm} >{\centering\arraybackslash}m{2.8cm}}
\toprule
\multirow{4}{*}{Group} & \multirow{4}{*}{Metric} & \multicolumn{4}{c}{Submodel Combination (A and B)} \\ \cline{3-6}
& & I\\A: CNN \\B: CNN & II\\A: ConvLSTM\\B: ConvLSTM & III\\ A: ConvGRU\\B: ConvGRU& IV \newline A: CNN-Attn \newline B: BERT\\ 
\midrule\midrule
\multirow{4}{*}{1} & Weighted-AUC & 0.929$\pm$0.004 & 0.860$\pm$0.006 & 0.917$\pm$0.004 & 0.817$\pm$0.004 \\
                   & F1  & 0.664$\pm$0.006 & 0.477$\pm$0.010 & 0.604$\pm$0.006 & 0.442$\pm$0.007 \\
                   & AUP-RC & 0.795$\pm$0.005 & 0.596$\pm$0.007 & 0.732$\pm$0.006 & 0.609$\pm$0.007 \\
                   & Elapsed Time &\textbf{ 17:14} & 33:09 & 30:54 & 5:41:06 \\
\hline
\multirow{4}{*}{2} & Weighted-AUC & 0.900$\pm$0.003 & 0.834$\pm$0.003 & 0.875$\pm$0.004 & 0.886$\pm$0.003 \\
                   & F1  & 0.587$\pm$0.009 & 0.475$\pm$0.006 & 0.566$\pm$0.010 & 0.524$\pm$0.007 \\
                   & AUP-RC & 0.650$\pm$0.003 & 0.527$\pm$0.004 & 0.650$\pm$0.004 & 0.635$\pm$0.009 \\
                   & Elapsed Time & 24:15 & 26:38 & 24:33 & 7:54:33 \\
\hline
\multirow{4}{*}{3} & AUC & \textbf{0.932$\pm$0.004} & 0.793$\pm$0.008 & 0.857$\pm$0.003 & 0.774$\pm$0.006 \\
                   & F1  & \textbf{0.678$\pm$0.011} & 0.366$\pm$0.010 & 0.509$\pm$0.008 & 0.355$\pm$0.010 \\
                   & AUPRC & \textbf{0.805$\pm$0.003} & 0.471$\pm$0.005 & 0.683$\pm$0.009 & 0.448$\pm$0.004 \\
                   & Elapsed Time & 17:54 & 23:37 & 27:16 & 6:14:33 \\
\hline
\multirow{4}{*}{4} & AUC & 0.867$\pm$0.003 & 0.830$\pm$0.008 & 0.887$\pm$0.004 & 0.847$\pm$0.004 \\
                   & F1  & 0.568$\pm$0.009 & 0.453$\pm$0.011 & 0.547$\pm$0.006 & 0.445$\pm$0.008 \\
                   & AUPRC & 0.618$\pm$0.003 & 0.502$\pm$0.009 & 0.718$\pm$0.003 & 0.593$\pm$0.01 \\
                   & Elapsed Time & 25:55 & 27:05 & 31:28 & 8:14:33 \\
\hline
\end{tabular}
}
\end{center}
\end{table*}

\subsection{Robustness Test}

We conduct two robustness evaluations: out-of-time (OOT) and out-of-universe (OOU) \cite{{mushava2024comprehensive}} to ensure the practicality of our model for real-world use. These evaluations can play a crucial role in analyzing the reliability and generalizability of our model to handle unseen data in real-world scenarios. We evaluate the OOT performance of the model by splitting the train and test data based on the time index. In OOT, we simulate future performance by testing the model on data from a later time period than the training data, ensuring it can be generalized to future observations. In OOU, we split the train and test data based on corporate entities, ensuring that the model is tested on entirely new corporates that were not present in the training set. This way, we assess the model's ability to generalize across different corporate entities, making it more reliable for application to new or previously unseen corporates in real-world settings.

We further validate the results using a 90 percent bootstrap confidence interval using 10,000 resamples. This approach allows us to assess the stability of the metric by estimating its distribution scores across different random samples from the test data, providing a robust measure of the model’s reliability \cite{diciccio1996bootstrap}. 
The results in Table~\ref{tab:ootoou} indicate a high level of performance, with a weighted AUC score of 0.935 for the time-based test data, indicating that the model’s performance is robust even under subsampling.

Furthermore, we assess the OOU by testing the model on a dataset containing companies that were not present in the training set. The results in Table~\ref{tab:ootoou} show a high level of OOU performance, with a weighted AUC score of 0.929, indicating that the model could accurately predict both previously seen and unseen companies. Furthermore, the patterns found were structural rather than company-specific. 

The results imply that the model can identify general relationships between various data modalities, rather than relying on unique characteristics of specific companies in the training set. The model can generalize effectively to new companies, as it has learned how both structured numerical data and unstructured textual data influence credit ratings in a broad, cross-company context.
 
\begin{table}[!h]
\begin{center}
\caption{Robustness test for the best model showing OOT and OOU and comparing them with the best model performance as a baseline. A 90 percent bootstrap confidence interval (CI) was calculated using 10,000 resamples to provide a robust estimate of the model's performance stability.}
\label{tab:ootoou}
\begin{tabular}{cccc}
\toprule
Metric& OOT& OOU& Baseline \\\midrule\midrule
Weighted-AUC &0.935$\pm$0.004&0.929$\pm$0.004& 0.932$\pm$0.004\\
F1 & 0.689$\pm$0.011& 0.66$\pm$0.011& 0.678$\pm$0.011\\\bottomrule
\end{tabular}
\end{center}
\end{table}

\subsection{The role of data modality in prediction}

We evaluated the role of data modalities in the prediction and investigated which modality or data channel (text or numerical) had contributed the most using the best model (Group 3 - Combination I in Figure \ref{fig:group3comb1}). 

 
 We retrain the best model using each channel individually as input by separating the original dataset into different subsets, where each subset contains only one type of data (e.g., text data, financial ratios, bond data, or market data). This allows us to isolate the contribution of each data channel. In each subset, we train the best model separately (by removing the data input from other channels) to evaluate how each data type affects the prediction. Once the model is trained on each subset, we evaluate its performance on the test set and compare the classification accuracy. Additionally, we calculate a 90 percent bootstrap confidence interval using 10,000 re-samples to provide a robust estimate of the performance stability for each channel.

Tables \ref{tab:textchan} and \ref{tab:numericalchan} show that the text channel had the most contribution (weighted AUC of 0.915), which was significantly higher than the numerical channel. In the numerical channel, the highest contribution was related to covariate and financial ratio channels with weighted AUC values of 0.808 and 0.791, respectively, suggesting that the information contained in these channels had been more informative for prediction than that in the other unstructured channels. The covariate channel provided information regarding the rating agency type and the last observed rating class; this information may have been crucial because the rating agency type can affect the reliability and accuracy of the ratings, and the last observed rating class can provide insights into the historical creditworthiness of the company. The financial ratio channel provided a wide range of features related to the financial performance of the companies and could provide insights into their financial health and their ability to repay debts.

\begin{table}[!h]
\begin{center}
\caption{Performance of text channel for the best model, showing the Weighted-AUC and F1-score. A 90 percent bootstrap confidence interval (CI) was calculated using 10,000 resamples.}
\label{tab:textchan}
\begin{tabular}{cc}
\toprule
Metric&	Text contribution \\\midrule\midrule 
Weighted-AUC&0.915 $\pm$0.004\\
 F1-score	&  0.654$\pm$0.011\\\bottomrule 
\end{tabular}
\end{center}
\end{table}

\begin{table}[!h]
\begin{center}
\caption{Performance of numerical channel for the best model, showing the Weighted-AUC and F1-score. A 90 percent bootstrap confidence interval (CI) was calculated using 10,000 resamples}
\label{tab:numericalchan}
\begin{footnotesize}
\begin{tabular}{ccccc}
\toprule
Metric&Covariate&	Fin. Ratio	&Bond&	Market \\\midrule\midrule
Weighted-AUC&0.808$\pm$0.006&0.791$\pm$0.006 &0.777$\pm$0.006 &0.53$\pm$0.007\\
F1-score&0.444$\pm$0.011&0.35$\pm$0.011	& 0.339$\pm$0.011	& 0.095$\pm$0.006\\\bottomrule
\end{tabular}
\end{footnotesize}
\end{center}
\end{table} 

\subsection{The effect of COVID-19 on model performance}


 The COVID-19 crisis significantly impacted the global economy, including financial markets, which affected various features (such as financial ratios and market indicators) useful for building prediction models. In this analysis, we examined the effect of the COVID-19 pandemic on credit rating prediction using the best-performing model (Group 3 - Combination I in Figure \ref{fig:group3comb1}). The timeline for the data used in training and testing is divided into two distinct periods. Both the training and test datasets for the pre-COVID period cover January 2010 until the start of the pandemic in early 2020. In the post-COVID analysis, both the training and test datasets span from the beginning of the pandemic in early 2020 through December 2022. Table~\ref{tab:covid} shows the model’s performance for both the pre-COVID and post-COVID periods, focusing on the test datasets from each period for comparative analysis. We find that the weighted AUC value increased from 0.927 to 0.964, suggesting that the model’s performance was improved. This improvement may have occurred because many predictable trends in the data became less credible after the crisis. The numerical data that was previously useful in prediction models became less relevant and less reliable. This effect led to a shift in the importance of information, as executives could provide useful context to the challenges or competitive advantage each company had in the face of the pandemic, which became more significant in affecting corporate credit ratings.

\begin{table}[htbp!]
\begin{center}
\caption{Comparison of the model before COVID-19 and  during and after COVID-19, showing Weighted-AUC and F1-score and 90\% bootstrap confidence intervals.} 
\label{tab:covid}
\begin{tabular}{ccc}
\toprule
Metric&	Before COVID-19&	During/After  COVID-19 \\\midrule\midrule
Weighted-AUC&0.927$\pm$0.004&0.964$\pm$0.007 \\
 F1-score	& 0.657$\pm$0.011&	0.777$\pm$0.026\\\bottomrule
\end{tabular}
\end{center}
\end{table} 
 
\subsection{Effect of the rating  agency on the time-lag of prediction}

\begin{figure}[htbp!]
\centering
\includegraphics[width=3in]{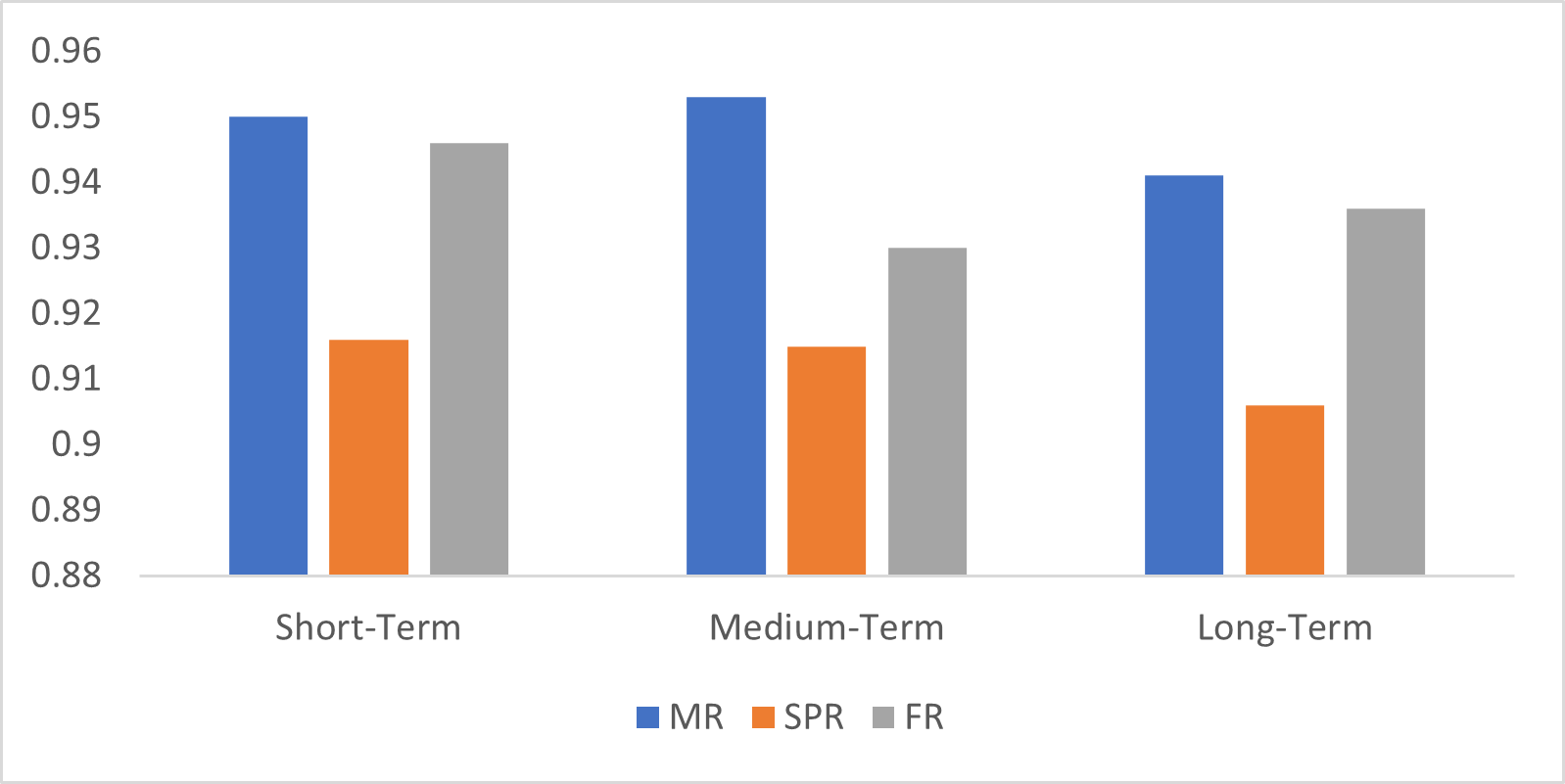}
\caption{Performance by prediction term and rating agency type (Moody's (MR), Standard \& Poor's (SPR), and Fitch Ratings (FR)), showing short-term (up to 4 months), medium-term (5 to 9 months), and long-term (more than 9 months) forecast accuracy. The y-axis represents the Weighted-AUC. }
\label{fig:lag}
\end{figure}

We further investigate the impact of agency rating type (e.g. Moody's) on the time-lag of prediction both simultaneously and separately using the best  model (Group 3 - Combination I in Figure \ref{fig:group3comb1}). The time-lag of prediction refers to the time elapsed between making a prediction and observation of the actual outcome. We present the results evaluating the significance of different agency types including Moody’s Rating  (MR)\footnote{\url{https://www.moodys.com/}}, Standard $\&$ Poor's Rating (SPR)\footnote{\url{https://www.spglobal.com/ratings/en/}} and Fitch Rating (FR)\footnote{\url{https://www.fitchratings.com/}} \cite{white2010markets} and time-lags (short-term, medium-term and long-term) as shown in  Table~\ref{tab:ajency} and Figure~\ref{fig:lag}.

\begin{table*}[htbp!]
\begin{center}
\caption{AUC and F1 score by Agency Type \& Terms}
\label{tab:ajency}
\adjustbox{max width=1\textwidth}{%
\begin{tabular}{>{\centering}m{0.7cm}>{\centering}m{2.2cm}>{\centering}m{2.2cm}>{\centering}m{2.2cm}>{\centering\arraybackslash}m{2.2cm}}
\toprule
Type&		All	&Short-Term&	Medium-Term& 	Long-Term \\\midrule\midrule
All &0.933$\pm$0.003 &0.936$\pm$0.006&0.935$\pm$0.005& 	0.925$\pm$0.008\\
&0.668$\pm$0.01	& 0.691$\pm$0.018&	 0.672$\pm$0.015& 0.643$\pm$0.022 \\\midrule
MR	 &0.945$\pm$0.005&0.945$\pm$0.009&0.951$\pm$0.006&0.938$\pm$0.012\\
	& 0.719$\pm$0.015	& 0.719$\pm$0.026	& 0.702$\pm$0.022&	 0.652$\pm$0.035\\\midrule
 SPR	&0.909$\pm$0.008&0.913$\pm$0.013&0.911$\pm$0.011&0.898$\pm$0.016 \\
 	&  0.604$\pm$0.018	&  0.609$\pm$0.034	& 0.597$\pm$0.027&	0.595$\pm$0.036 \\\midrule
 FR	&0.938$\pm$0.008&0.944$\pm$0.013&	0.933$\pm$0.012&0.941$\pm$0.015\\
 & 	 0.687$\pm$0.024	&  0.694$\pm$0.043&  0.683$\pm$0.033	& 0.674$\pm$0.045\\\bottomrule
\end{tabular}}
\end{center}
\end{table*} 

The first column of Table~\ref{tab:ajency}indicates the impact of rating type without considering lags. We observe that the rating performance based on MR is the most accurate one, followed by FR and SPR. This shows that for our specific dataset in midcaps, Moody’s provided the most accurate prediction of the point-in-time credit risk for these companies. As each rating was composed both of quantitative and qualitative information, we can infer that Moody’s and Fitch followed the market more closely than S\&P, which could either be focusing on through-the-cycle ratings or somehow missing the prediction of company collapse or 'default'. Nevertheless, all the rating companies demonstrated strong performance overall.


The short term refers to the prediction of rating in less than 4 months, and the medium term refers to 5 to 9 months. Finally, long-term prediction refers to predictions of more than 9 months. As expected, the short-term prediction outperformed the median and long-term ones (because credit risk is generally more predictable over shorter time horizons than longer ones), but this difference was very minor. In general, 1-year look ahead periods are very achievable with deep learning models.


 We finally analyze the results by combining the effect of the rating agency type with the prediction time horizon (as shown in the second row of Table~\ref{tab:ajency}). We notice that the medium-term predictions based on Moody's (MR) ratings  outperformed the short-term predictions; however, we expected that the short-term predictions would perform better due to the smaller time window. This suggests that Moody’s rating was more accurate in forecasting credit risk over a slightly longer period (5 to 9 months), possibly due to the nature of the data and the type of analysis used by the agency for their own credit-rating system. This difference is  relatively minor, and the contradiction may have arisen because the text channel contributes more than the numerical channel  for   credit rating prediction. Thus, it would not seem unrealistic that the impact of the transcript would become more apparent over a longer period, depending on the agency-type rating of the target data. 

\section{Discussion}
Our study demonstrated the capabilities of multimodal deep learning to improve credit rating predictions by utilizing both structured and unstructured data. Our framework not only overcomes the constraints of traditional unimodal methods but also advances the frontier of predictive analytics in financial environments. We explored how a combination of deep learning-based submodels can effectively handle complex data interactions through the employment of diverse data fusion strategies.

The comparative analysis of fusion techniques underscores the nuanced impact of data integration methods on model performance. Our results demonstrated that simple concatenation, while less complex, often fell short of achieving the depth of interaction necessary between modalities compared to cross-attention mechanisms which facilitated a more dynamic and context-aware amalgamation of features (Table \ref{tab:resultscomb}. This finding suggests that the complexity of data relationships in credit rating scenarios is better captured through sophisticated interaction models, which can discern the relevance and influence of different data streams more effectively.
Moreover, the study highlights the critical role of unstructured data, particularly textual information from earnings calls, in refining the predictions. The textual data contributed substantially to the models' performance, indicating that qualitative insights extracted from unstructured formats are pivotal in understanding and predicting financial behaviours and outcomes. This aligns with emerging trends in financial analytics, where qualitative data is increasingly recognized for its depth of context and predictive value.

The robustness tests (OOT and OOU evaluations, Table~\ref{tab:ootoou}), further reinforce the model's applicability and reliability in real-world settings. These tests ensure that the model is not only theoretically sound but also practical and resilient to changes in data patterns over time. The adaptability of the model to the disruptions caused by the COVID-19 pandemic particularly underscores its capability to handle external shocks (\ref{tab:covid}, making it a valuable tool for financial institutions that require resilience in their predictive models.
The BERT model was less effective across configurations due to its token limitation of 512, which led to truncation in processing longer text sequences (average cleaned token length = 4,000) (elapsed time in Table \ref{tab:resultscomb}. Other transformer-based approaches can be used to account for longer contexts, although the data size will remain an issue.  The other models (CNN, ConvLSTM and ConvGRU) did not face this limitation, allowing them to process the full information faster and perform better. CNN-based  submodels, in particular, required less training time since they featured fewer parameters when compared to CNN-Atten submodel.

Although there is no specific state-of-the-art (SOTA) in the literature on multimodal deep learning for credit rating prediction, several advancements in financial applications of deep learning provide valuable comparisons. For example, Cheng et al. \cite{cheng2024multi} and Wanliu Che et al. \cite{che2024predicting} explored complex multimodal data fusion techniques, which resonate with our framework, but have been used in different contexts such as call centres and financial distress prediction, respectively. Similarly, Zhang et al. \cite { zhang2024smpdf} integrated structured and unstructured data to enhance stock market predictions, demonstrating the effectiveness of techniques that could be adapted for credit rating prediction. These models underline the growing sophistication in handling multimodal data and the challenges related to computational demands and data quality dependence. 
There are several related papers in this area that have also studied the use of deep learning models for credit rating prediction\cite{yijun2009artificial,feng2020every,golbayani2020application,wang2015survey,tsai2010credit}. However, our study differs in that it evaluates a wide range of multimodal models and fusion type-and-level strategies, providing a more comprehensive evaluation of different approaches.

\subsection{Limitation of the Study}

A significant limitation of our current models for credit rating prediction is their reliance on the availability and quality of financial data, including earning call reports. These models assume that such reports are available for all companies being rated, which is not always the case, particularly for private companies. Additionally, the performance of these models is sensitive to the nature and quality of the data they process. They depend heavily on historical financial data and external market conditions, making them vulnerable to changes in data collection processes or shifts in economic environments. Several strategies can be employed to improve the generalizability and robustness of our credit rating framework. First, diversifying the sources of data used in the models can reduce reliance on specific types of financial reports and broaden the scope of data analysis. Implementing techniques such as cross-validation can help mitigate the risk of overfitting by ensuring the models perform well across various subsets of data. Additionally, developing adaptive models that can adjust to changes in economic conditions and data quality may enhance performance stability.

Another limitation is the lack of uncertainty quantification in model predictions, which refers to the estimation of the degree of uncertainty associated with the predictions made by the model. The framework can utilize submodels that use Bayesian deep learning \cite{chandra2021revisiting, wang2020survey}; however, they also have challenges for models with larges number of parameters.

Another limitation arises from the diverse standards employed by different rating agencies. Each agency may use different criteria or emphasize various aspects of creditworthiness differently, leading to discrepancies in the credit rating. To address this issue in our study, we used data from three different agencies and employed an equivalency table to standardize ratings across these agencies. However, despite these efforts, some inconsistencies and methodological differences remain that could still impact the generalizability and accuracy of our models.  A strategy for improvement can be incorporating a more sophisticated normalization or translation framework that accounts for the unique rating philosophies of each agency could help in reducing discrepancies. This might involve using advanced machine learning techniques such as explainable artificial intelligence (xAI)  \cite{saeed2023explainable} to learn and understand mappings between the ratings of different agencies.

\subsection{Future Work}
Improving the performance of the multimodal framework can involve innovative strategies, focusing on diverse layer architectures, novel fusion models, and different levels of fusion.  
Additionally, we can explore graph-based deep learning models to enhance sequential data processing and modality relationships \cite{wu2020comprehensive}. We can also use advanced fusion techniques like probabilistic fusion models incorporating Bayesian models for uncertainty quantification \cite{siahkoohi2020deep} and multiplicative integration \cite{jayakumar2020multiplicative}  to improve uncertainty estimation and feature interaction. Furthermore, it explores various fusion levels, including late fusion for independent modal strengths, hierarchical fusion for multi-level feature interactions, and adaptive fusion to allow the model to self-adjust its fusion strategy based on specific tasks, potentially enhancing overall model robustness and performance.
Additionally, incorporating causal inference techniques to examine the relationships between extracted features and creditworthiness could yield more actionable insights. This approach would not only identify patterns and correlations but also uncover the underlying causal mechanisms influencing creditworthiness, enhancing both model transparency and decision-making effectiveness.

Our model excels in domains that necessitate the integration of multimodal data due to its ability to effectively process and analyze diverse types of information simultaneously. For instance, in image processing, it effectively combines image data with text and sensor inputs to enhance medical diagnostics, improving disease detection and diagnostic efficacy. In bioinformatics, the model integrates genetic sequences, structures, and functional data, expediting the prediction of protein functions and gene-disease associations, which can revolutionize personalized medicine by identifying key biomarkers and therapeutic targets. For sentiment analysis, it merges textual content with demographic and contextual details, enhancing the analysis of social media sentiments and refining sentiment detection for more precise market research and public opinion analysis.

Our framework's flexibility in handling diverse financial data types makes it suitable for a range of financial sectors. Its ability to process and integrate multiple data modalities—including text, numerical, and even other unstructured data—is critical for diverse applications such as portfolio management, where it is necessary to synthesize vast market and asset information; insurance risk assessment, which demands analysis of varying risk factors and client data; and investment analysis, where integrating historical financial data with real-time market feeds enhances the insights into potential risks and opportunities. An additional application is country-based credit rating prediction \cite{da2019sovereign}, where the model utilizes economic, political, and financial indicators to assess the creditworthiness of nations, aiding governments and investors in making informed decisions. These capabilities are further extendable through specific adaptations to meet unique regulatory and operational requirements across different financial domains. Our framework can be altered to support dynamic adjustments to changing market conditions through its capability to seamlessly integrate real-time data updates. Enhancing the model’s architecture to include adaptive learning techniques such as online learning allows for continual model updates without full retraining. This is ideal for environments with constant data flow, such as high-frequency trading or real-time credit risk assessment. Additionally, implementing reinforcement learning enables the model to optimize decision-making by learning from the outcomes of past decisions, thereby aligning with business objectives such as profit maximization or risk minimization.

The robust and versatile nature of our framework provides a solid foundation for interdisciplinary applications beyond the financial sector. Potential fields include healthcare, where it could improve diagnostic accuracy by integrating medical data types, and environmental science, predicting natural disasters or pollution levels through diverse data sources. Such applications demonstrate the model's potential to significantly impact areas like urban planning and social sciences, aiding in resource allocation and policymaking.

\section{Conclusions}
\label{sec:conclusions}
 
In this study, we presented a framework for the evaluation of combination of multimodal models consisting of data fusion level strategies, combining deep learning models for credit rating prediction that featured text and numerical data channels. We have shown that a CNN-based model with early-intermediate hybrid fusion with simple concatenation provided the best accuracy for our credit rating problem.  
In addition, we conducted a test of robustness to ensure that these models would be applicable when provided with newer information, contributing to the stability of our model. Furthermore, analysis of the contribution of each modality to the prediction showed that the text channel played the most significant role, showcasing the importance of the context, particularly during the COVID-19 pandemic. Finally, we investigated the impact of the type of rating agency on  the time lag and discovered that the credit rating by Moody’s was more accurate, suggesting a point-in-time model, than the others, particularly in the medium term. The results of our study can be used by rating agencies and other financial institutions to make more informed decisions regarding credit ratings.In addition, the results regarding the reliability of the model during unexpected crises can be useful in terms of developing more robust and reliable models during such times.
 
Our study  not only advances our understanding of multimodal data fusion in credit rating prediction, but also sets the stage for future explorations into more complex and holistic approaches to financial modelling. The insights gained from this study could potentially guide the development of next-generation financial analytics tools that are accurate,  and robust, thereby supporting better decision-making in the financial industry.

\section{Code}
The source code of the framework can be found at the following GitHub repository
\footnote{\url{https://github.com/Banking-Analytics-Lab/MultimodalFusionRatings}}. 

 \section*{Acknowledgment}
M. Tavakoli, F. Tian and C. Bravo acknowledge the support of the NSERC Discovery Grant program [RGPIN-2020-07114]. This research was partly funded by the Canada Research Chairs program [CRC-2018-00082] and by Compute Ontario (\url{computeontario.ca}), Calcul Québec (\url{calculquebec.ca}), and the Digital Research Alliance of Canada (\url{alliancecan.ca}).

\appendix
\section{Notation Summary}\label{ref:NotationApp}
Below is a summary of notations used throughout the manuscript, organized into different categories for clarity.

{\footnotesize
\centering
\begin{longtable}{>{\bfseries}l p{12cm}}
\caption{Summary of Notations}
\\
\toprule
\textbf{Symbol} & \textbf{Description} \\
\midrule
\multicolumn{2}{c}{\textbf{Model Components and Operations}} \\
\midrule
A, B & Labels representing distinct network architectures in the models. Network A for numerical data, Network B for text. \\
Conv & Convolutional Layer, used for feature extraction from input data. \\
MaxP & Max Pooling Layer, reduces the dimensionality by selecting the maximum value within a pooling window. \\
Drop & Dropout Layer, randomly drops units during training to prevent overfitting. \\
GlobAve & Global Average Pooling Layer, reduces the entire feature map to a single value by averaging. \\
LSTM & Long Short-Term Memory Layer, a recurrent neural network layer for sequence prediction. \\
GRU & Gated Recurrent Unit Layer, a simplified variant of LSTM for sequence modeling. \\
ConvLSTM & A combination of Convolutional and Long Short-Term Memory layers, useful for spatiotemporal data. \\
ConvGRU & A combination of Convolutional and Gated Recurrent Unit layers, useful for sequence prediction in spatiotemporal data. \\
CNN-Attn & Convolutional Neural Network combined with an integrated Attention mechanism to focus on relevant features within the data. \\
BERT & Bidirectional Encoder Representations from Transformers, a pre-trained model used for text processing. \\
\midrule
\multicolumn{2}{c}{\textbf{Data Types and Sources}} \\
\midrule
Bond & Historical data related to bond securities, including features like price, volume, etc. \\
Market & Market data, including indices and market capitalization. \\
Fin Ratio & Financial ratios of companies indicating profitability, liquidity, etc. \\
Covariate & Data containing covariates such as rating agency type and previous ratings. \\
Text & Unstructured text data, such as earnings call transcripts. \\
\midrule 
\multicolumn{2}{c}{\textbf{Model Metrics}} \\
\midrule
AUC & Area Under the Curve measures the ability of the model to classify correctly across different thresholds. \\
F-1 Score & Harmonic mean of precision and recall, used as a metric for binary classification performance. \\
AUP-RC & Area Under the Precision-Recall Curve, measures the trade-off between precision and recall across different thresholds, especially useful for imbalanced datasets. \\
\bottomrule
\end{longtable}
}
\end{document}